\documentclass[lettersize,journal]{IEEEtran}
\usepackage{amsmath,amsfonts}
\usepackage{algorithmic}
\usepackage{array}
\usepackage[caption=false,font=normalsize,labelfont=sf,textfont=sf]{subfig}
\usepackage{textcomp}
\usepackage{stfloats}
\usepackage{verbatim}
\usepackage{graphicx}
\def\BibTeX{{\rm B\kern-.05em{\sc i\kern-.025em b}\kern-.08em
    T\kern-.1667em\lower.7ex\hbox{E}\kern-.125emX}}
\usepackage{balance} 
\usepackage{booktabs}
\usepackage{makecell}
\usepackage{multirow}
\usepackage{color}
\usepackage{floatrow}
\newfloatcommand{capbtabbox}{table}[][\FBwidth]
\usepackage{etoolbox}
\usepackage[ruled,linesnumbered]{algorithm2e}
\makeatletter
\patchcmd{\@makecaption}
  {\scshape}
  {}
  {}
  {}
\makeatother

\usepackage{hyperref}
\makeatletter
\def\UrlAlphabet{%
      \do\a\do\b\do\c\do\d\do\e\do\f\do\g\do\h\do\i\do\j%
      \do\k\do\l\do\m\do\n\do\o\do\p\do\q\do\r\do\s\do\t%
      \do\u\do\v\do\w\do\x\do\y\do\z\do\A\do\B\do\C\do\D%
      \do\E\do\F\do\G\do\H\do\I\do\J\do\K\do\L\do\M\do\N%
      \do\O\do\P\do\Q\do\R\do\S\do\T\do\U\do\V\do\W\do\X%
      \do\Y\do\Z}
\def\UrlDigits{\do\1\do\2\do\3\do\4\do\5\do\6\do\7\do\8\do\9\do\0}
\g@addto@macro{\UrlBreaks}{\UrlOrds}
\g@addto@macro{\UrlBreaks}{\UrlAlphabet}
\g@addto@macro{\UrlBreaks}{\UrlDigits}
\makeatother
\begin{document}
\title{A Novel RFID Authentication Protocol Based on A Block-Order-Modulus Variable Matrix Encryption Algorithm}
\author{Yan Wang, Ruiqi Liu,~\emph{Senior Member},~\emph{IEEE}, Tong Gao, Feng Shu, Xuemei Lei, Yongpeng Wu, Guan Gui,~\emph{Fellow},~\emph{IEEE}, Jiangzhou Wang,~\emph{Fellow},~\emph{IEEE}
\thanks{Manuscript created October, 2020; This work was supported in part by the National Natural Science Foundation of China under Grant U22A2002, and  by the Hainan Province Science and Technology Special Fund under Grant ZDYF2024GXJS292; in part by the Scientific Research Fund Project of Hainan University under Grant KYQD(ZR)-21008; in part by the Collaborative Innovation Center of Information Technology, Hainan University, under Grant XTCX2022XXC07; in part by the National Key Research and Development Program of China under Grant 2023YFF0612900. 
(Corresponding author: Feng Shu).}
\thanks{Yan Wang is with the School of Information and Communication Engineering, Hainan University, Haikou 570228, China (e-mail: yanwang@hainanu.edu.cn).}
\thanks{Ruiqi Liu is with the Wireless and Computing Research Institute, ZTE Corporation, Beijing 100029, China (e-mail: richie.leo@zte.com.cn).}
\thanks{Tong Gao is with the College of Electronic Science and Engineering, Jilin University, Changchun 130012, China (e-mail: gaotong@jlu.edu.cn).}
\thanks{Feng Shu is with the School of Information and Communication Engineering and Collaborative Innovation Center of Information Technology, Hainan University, Haikou 570228, China, and also with the School of Electronic and Optical Engineering, Nanjing University of Science and Technology, Nanjing 210094, China (e-mail: shufeng0101@163.com).}
\thanks{Xuemei Lei is with the College of Electronic Information Engineering, Inner Mongolia University, Hohhot 010021, China (e-mail: ndlxm@imu.edu.cn).}
\thanks{Yongpeng Wu is with the Shanghai Key Laboratory of Navigation and
Location Based Services, Shanghai Jiao Tong University, Minhang, Shanghai,
200240, China (e-mail:  yongpeng.wu2016@gmail.com).}
\thanks{Guan Gui is with the College of Telecommunications and Information Engineering, Nanjing University of Posts and Telecommunications, Nanjing 210003, China (e-mail: guiguan@njupt.edu.cn).}
\thanks{Jiangzhou Wang is with the School of Engineering, University of Kent, CT2 7NT Canterbury, U.K. (e-mail: j.z.wang@kent.ac.uk).}}


\maketitle

\begin{abstract}
In this paper, authentication for mobile radio frequency identification (RFID) systems with low-cost tags is studied. Firstly, an adaptive modulus (AM) encryption algorithm is proposed. Subsequently, in order to enhance the security without additional storage of new key matrices, a self-updating encryption order (SUEO) algorithm is designed. Furthermore, a diagonal block local transpose key matrix (DBLTKM) encryption algorithm is presented, which effectively expands the feasible domain of the key space. Based on the above three algorithms, a novel joint AM-SUEO-DBLTKM encryption algorithm is constructed. Making full use of the advantages of the proposed joint algorithm, a two-way RFID authentication protocol, named AM-SUEO-DBLTKM-RFID, is proposed for mobile RFID systems. In addition, the Burrows-Abadi-Needham (BAN) logic and security analysis indicate that the proposed AM-SUEO-DBLTKM-RFID protocol can effectively combat various typical attacks. Numerical results demonstrate that the proposed AM-SUEO-DBLTKM algorithm can save 99.59\% of tag storage over traditional algorithms. Finally, the low computational complexity as well as the low storage cost of the proposed AM-SUEO-DBLTKM-RFID protocol facilitates deployment within low-cost RFID tags.
\end{abstract}


\section{Introduction}
\IEEEPARstart{T}{he} 6th generation (6G) wireless technologies are envisaged to support and empower vertical industries \cite{6G}, including the ones enabled by massive internet of things (IoT) devices.
Thanks to their continuous advancement, IoT devices have been employed widely in various industries including manufacturing, logistics, smart healthcare, and intelligent cities \cite{KIoT-QWatch}. 
As a critical technology to implement IoT, radio frequency identification (RFID) has been extensively applied due to its unique advantages of non-contact and simultaneous recognition of multiple objects \cite{HNew}.
For example, the implement of RFID technology enables efficient information management within the logistics procurement chain, thereby boosting the effectiveness and precision of logistics distribution. \cite{FBroadband/Dual-Band}.

On the other hand, with the rapid deployment of RFID in a multitude fields, its security and privacy issues have also emerged.
The authors of \cite{GPhysical-Layer} pointed out that due to low computational capabilities, the chip-less sensory tags were unable to adopt mature and complex encryption mechanisms to protect themselves.
Consequently, low-cost RFID tags are currently vulnerable to diverse attacks, including denial of service (DoS) \cite{FPUF}, impersonation attack \cite{SundaresanA}, de-synchronization \cite{GopeLightweight}, man-in-the-middle attacks \cite{ZhangDevice-Side}, or replay attacks \cite{QiSAE}.
These attacks may greatly impede the further application of RFID technologies.
For instance, in RFID enabled smart healthcare systems, improper authentication protocols may result in the failure to provide accurate information promptly, causing medical professionals to make erroneous treatment decisions and possibly endanger the well beings of patients \cite{DAn}.
Furthermore, the authors of \cite{KLightweight} mentioned that personal medical confidential information can be disclosed to insurance firms, which not only violated the privacy entitlements of people, but also impeded the harmonious progress of the medical services sector.
Moreover, if the data of RFID tags is maliciously tampered with, it may lead to goods being delivered to the wrong place, causing delays or even paralysis in the supply chain system \cite{CPrivacy}.
In summary, in order to make RFID technologies better serve people's lives, it is desirable to effectively tackle the security and privacy challenges.

It is reassuring to know that the security issues of RFID technologies have been paid much attention and have been already proposed its solutions for different use cases.
For instance, a comprehensive set of guidelines for RFID security and privacy, backed by modelling principles, mitigation strategies, and attack vectors were proposed by the authors of \cite{HAnalysis}.
In 2018, the authors of \cite{SA} proposed a robust authentication protocol based on elliptic curve cryptography (ECC) for telecare medical information systems (TMIS). 
Nevertheless, the protocol was unable to safeguard against replay attacks, preserve client anonymity, thwart impersonation attacks, or prevent password-guessing attacks.
Subsequently, a cloud-based authentication protocol that is both effective and dependable was presented in \cite{KCloud-based}, in which the authors incorporated bit-wise rotation, permutation, and public-key encryption techniques in this protocol to effectively defend against threats like tracking, de-synchronization, and replay attacks. 
Although this protocol provided higher-level security, it was not impractical for low-cost RFID tags because of its substantial computational requirements.
In addition, the authors of \cite{VRseap} constructed an authentication protocol based on timestamp and ECC. 
However, an efficacious impersonation attack on tags and readers was implemented by \cite{MRseap2} against this design. 
It is worth noting that the aforementioned RFID authentication protocols typically use mature and secure encryption algorithms, such as ECC. 
However, the computational abilities and storage facilities of low-cost RFID tags are limited, which may not meet the technical requirements of these encryption algorithms.

In order to cater to low-cost RFID tags, many lightweight protocols using relatively low complexity encryption algorithms have been proposed \cite{Yang2022Delegating}.
Firstly, two lightweight RFID mutual authentication schemes for IoT environment were shown in \cite{KFanLightweight}.
These protocols utilized a hash function and a random number (RN) generator for secure authentication, greatly reducing computational and transmission costs.
However, the analysis and demonstration conducted by the authors in \cite{CTTowards} revealed that it was unable to successfully carry out attacks including reader spoofing, tag falsification, and message interception.
In \cite{SDesigning}, a novel and effectual lightweight RFID-based authentication protocol, named lightweight blockchain-enabled RFID-based authentication protocol (LBRAPS), was devised specifically for supply and demand chains operating within a 5th generation (5G) mobile edge computing circumstance, incorporating blockchain technology.
LBRAPS relied solely on one-way cryptographic hashing, bitwise exclusive-or (XOR), and bitwise rotation operations, resulting in reduced computational overhead.
However, the LBRAPS protocol had been analyzed to lack forward security \cite{SDrone}.

\begin{figure}[h]
\centering
	\includegraphics [width=0.8\textwidth]{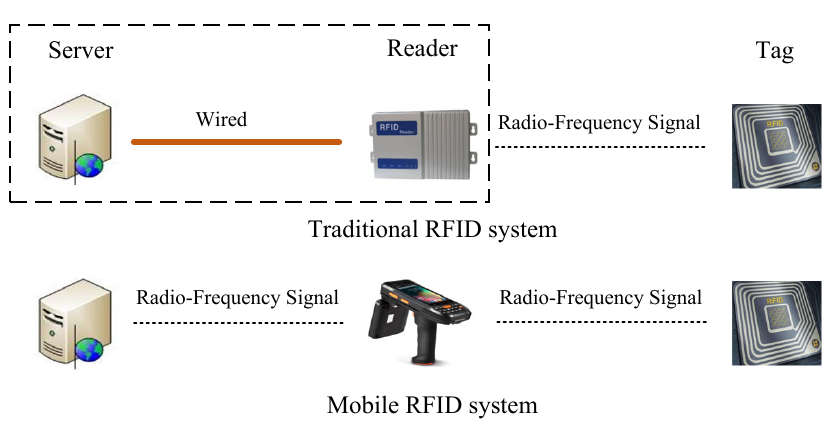}
	\caption{Comparison between a traditional RFID system and a mobile RFID system.}
    \label{Traditional-mobile}
\end{figure}

From the above analysis, it can be seen that the mature cryptographic algorithms pose challenges to the storage space and computing power of low-cost RFID tags, while the lightweight protocols, although meeting the requirements of low computational overhead, have relatively weak security.
Therefore, in order to have better trade-off between security, computational cost, tag storage, and other aspects, a RFID authentication protocol based on permutation matrix encryption was presented in \cite{FanPermutation}. 
This protocol based on matrix encryption algorithms had fast authentication speed and was suitable for low-cost RFID tags.
In 2022, the authors of \cite{YLuo} proposed a random rearrangement block matrix encryption algorithm. 
In the same year, a proficient RFID authentication scheme based on key matrix was introduced in \cite{WANGEfficient}.
However, these protocols have been proposed based on the assumption of secure communication between servers and readers, which are not always true in mobile RFID systems.
In practical applications, the application scenarios of traditional RFID systems with fixed reader positions are extremely limited. 
As depicted in Fig.~\ref{Traditional-mobile}, mobile RFID systems with handheld readers have high flexibility and are more popular in real-world implementation.

In summary, developing an authentication protocol that fulfills the security needs of mobile RFID systems while being feasible for deployment on low-cost RFID tags has emerged as a pivotal challenge.
Creating an RFID authentication protocol that strikes a balance between security and storage expenses is especially crucial in application contexts involving low-cost RFID tags with limited computing capabilities and resources.
In view of the above situation, the contributions of this paper can be outlined as follows.

\begin{enumerate}
  \item In traditional key matrix encryption algorithms, the constancy of the key matrix and modulus are easy to be stolen by attackers. In order to address this problem, firstly, an adaptive modulus (AM) encryption algorithm is proposed. Subsequently, to further boost security without additional storage of new key matrices, a self-updating encryption order (SUEO) algorithm is designed. In addition, inspired by the basic idea of matrix blocking and transposing, a novel diagonal block local transpose key matrix (DBLTKM) encryption algorithm is presented, which greatly extends the feasible domain of the key space.
  \item Based on the above three proposed encryption algorithms, their mixtures, namely AM-SUEO, AM-DBLTKM, SUEO-DBLTKM, AM-SUEO-DBLTKM encryption algorithm are proposed. In other words, a total of 7 key matrix encryption algorithms are developed for different application scenarios.  At the same time, the proof of feasibility of these encryption algorithms is provided. In addition, to further optimize the authentication efficiency of this joint algorithm, the fast convolutional Winograd algorithm is introduced. This significantly reduces the number of multiplication operations, further decreasing the computational complexity of the tag authentication process.
  \item In order to fully utilize the advantage that the proposed joint AM-SUEO-DBLTKM algorithm can improve security without increasing tag storage, a novel two-way RFID security authentication protocol is constructed based on this joint encryption algorithm. Subsequently, through the Burrows-Abadi-Needham (BAN) logic formal analysis and security analysis, it is shown that the proposed RFID authentication protocol can effectively resist various typical attacks. Finally, numerical results demonstrate that the novel RFID authentication protocol can save 99.59\% of tag storage over traditional algorithms. In terms of the computational complexity analysis of tags, it is shown that the protocol is friendly to be employed with low-cost RFID tags.
\end{enumerate}

The remainder of this paper is organized as follows. The mathematical modeling is presented in Section II. Subsequently, a novel joint AM-SUEO-DBLTKM encryption algorithm is proposed in Section III.
Furthermore, a novel two-way RFID authentication protocol based on the joint AM-SUEO-DBLTKM encryption algorithm is designed in Section IV. 
In Section V, the BAN logic, security analysis, and tag storage overhead of the newly proposed AM-SUEO-DBLTKM-RFID protocol are analyzed.
Numerical results as well as analysis are presented in Section VI, and finally, conclusions are drawn in Section VII.

Notations: Throughout the paper, boldface lower case and upper case letters represent vectors and matrices, respectively. 
The sign $\gcd(a, b)$ stands for the greatest common divisor of $a$ and $b$. 
The sign $\det(\cdot)$ denotes the determinant of a matrix. 
The sign $\mathbb{Z}$ represents the set of integers.
The sign $\mathbb{Z}^{+}$ stands for the set of positive integers.
The sign $\mathbb{Z}^{m\times n}$ denotes the set of $m\times n$ dimensional integer matrices.

\section{Mathematical modeling}

To begin with, the mathematical theorems involved in the key matrix encryption and decryption algorithm are given.

\textbf{\textit{Lemma 1:}} If $p$ is a positive integer, $a$ is an integer, and $p$ and $a$ are coprime, then the congruence equation $ax \equiv 1\bmod (p)$ has a unique solution in the sense of modulus $p$, i.e., there exists a positive integer $a^{\prime}<p$ make $aa^{\prime} \equiv 1\bmod (p)$.

\textbf{\textit{Corollary 1:}} If $\mathbf{A}, \mathbf{B}\in \mathbb{Z}^{n\times n}$, and $\mathbf{A}\times \mathbf{B}\equiv \mathbf{E} \bmod (p)$, where $\mathbf{E}$ is the identity matrix, then $\mathbf{B}$ is called the modulus $p$-inverse matrix of $\mathbf{A}$.

It is not difficult to prove that the condition for the existence of a modulus $p$-inverse matrix for $\mathbf{A}$ is that $\det(\mathbf{A})$ is coprime with $p$.
Therefore, the encryption and decryption process of the key matrix is as follows
\begin{align}\label{keymatrixE}
E(\mathbf{t},\mathbf{A},p)=\mathbf{A}\times \mathbf{t}\bmod{(p)}=\mathbf{c},
\end{align}
\begin{align}\label{keymatrixD}
D(\mathbf{c},\mathbf{B},p)=\mathbf{B}\times \mathbf{c}\bmod{(p)}=\mathbf{t},
\end{align}
where $\mathbf{t}$, $\mathbf{c}$, $\mathbf{A}$, $\mathbf{B}$, $E(\cdot)$, and $D(\cdot)$ represent plaintext vector, ciphertext vector, encryption matrix, decryption matrix, encryption process, and decryption process, respectively. 

According to the above encryption and decryption process, it can be seen that the immutability of the key matrix and the modulus $p$ is detrimental to security. In traditional RFID security authentication protocols based on key matrix, one possible method to improve security is to directly store multiple sets of encryption and decryption matrices. However, the disadvantage of this approach is that it will occupy a large amount of tag storage space, which is not conducive to deployment in low-cost RFID tags. Hence, how to boost the security of the algorithm without storing the new key matrix and modulus has become an urgent problem in the traditional key matrix encryption algorithm.

\section{Proposed a novel joint AM-SUEO-DBLTKM encryption algorithm}

\begin{figure}[h]
\centering
	\includegraphics [width=0.8\textwidth]{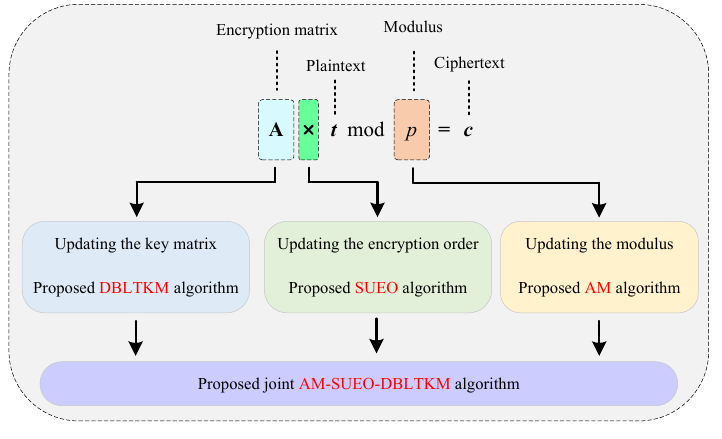}
	\caption{Different encryption algorithms are proposed from different perspectives and further joint encryption algorithms are presented.}
    \label{Section3zongtu}
\end{figure}

To elevate security without increasing the storage overhead of the tag, as shown in Fig.~\ref{Section3zongtu}, the following three aspects can be considered, i.e., updating the modulus, updating the encryption order, and updating the key matrix.
Correspondingly, the feasibility proofs of the proposed AM algorithm, SUEO algorithm and DBLTKM algorithm are presented in subsection A, subsection B and subsection C, respectively. Finally, a novel joint AM-SUEO-DBLTKM encryption algorithm is proposed in subsection D.

\subsection{Proposed AM encryption algorithm}

\textbf{\textit{Lemma 2:}} If positive integers $a$ and $p$ are coprime, then $a$ and $q$ are coprime, where $q$ is the integer divisor of $p$.

This lemma means that for an integer matrix $\mathbf{A}$, if $\det(\mathbf{A})$ and $p$ are coprime, then $\det(\mathbf{A})$ and the integer divisor $q$ of $p$ are also coprime. Based on \textbf{\textit{Corollary 1}}, it can be concluded that the $\mathbf{A}$ has a modulus $q$-inverse matrix.

\textbf{\textit{Corollary 2:}} If $\mathbf{A}, \mathbf{B}\in \mathbb{Z}^{n\times n}$, and $\mathbf{A}\times \mathbf{B}\equiv \mathbf{E} \bmod (p)$, where $\mathbf{E}$ is the identity matrix, then $\mathbf{B}$ is called the modulus $p$-inverse matrix of $\mathbf{A}$, and $\mathbf{B}$ is the modulus $q$-inverse matrix of $\mathbf{A}$, where $q$ is the integer divisor of $p$.

\textbf{\textit{Proof:}} Please refer to Appendix A.

\textbf{\textit{Lemma 3:}} $\forall a, b, c\in \mathbb{Z}^{+}$, if $\gcd(a, c)=1$ and $\gcd(b, c)=1$ exists, then $\gcd(a\times b, c)=1$.

This lemma shows that if $p$ is mutually prime with $\det(\mathbf{A})$ and $q$ is mutually prime with $\det(\mathbf{A})$, then $p \times q$ is still mutually prime with $\det(\mathbf{A})$.

According to the above proof, the encryption and decryption process of the key matrix can be achieved through modulus $p$, $q$ or $p \times q$, where $p$ is a composite number.
As shown in \textbf{Algorithm 1}, the total number of integer divisors (not containing 1) of modulus $p$ is $Q$. Therefore, the newly proposed AM algorithm extends the constant modulus to variable modulus, which can realize $Z_{\text{AM}}=2Q$ encryptions.

\begin{algorithm}[!h]
    \caption{Proposed AM encryption algorithm}
    \label{alg:AOA}
    \renewcommand{\algorithmicrequire}{\textbf{Input:}}
    \renewcommand{\algorithmicensure}{\textbf{Output:}}
    \begin{algorithmic}[1]
\REQUIRE plaintext $\mathbf{t}$, modulus $p$, secret value $S$, $N$ initial encryption matrices, AM index table
        \ENSURE ciphertext $\mathbf{c}$ 
        \STATE  Record the total number of elements of the AM index table as $Z_{\text{AM}}$;
        \STATE Calculate $m_1=\bmod(S,Z_{\text{AM}})$, and the value of $m_1$ determines which $q$ is selected for encryption, namely $m_1$ is the index value corresponding to the selected modulus $q$;
        \STATE Obtain $\mathbf{c}$ by (\ref{keymatrixE});
        \RETURN ciphertext $\mathbf{c}$.
    \end{algorithmic}
\end{algorithm}

\subsection{Proposed SUEO encryption algorithm}

\textbf{\textit{Lemma 4:}} If $\mathbf{A}, \mathbf{B}\in \mathbb{Z}^{n\times n}$, then $\det(\mathbf{A}\times\mathbf{B})
=\det(\mathbf{A})\times \det(\mathbf{B})$.

\textbf{\textit{Lemma 5:}} In general, if $\mathbf{A}, \mathbf{B}\in \mathbb{Z}^{n\times n}$, and $\mathbf{A}\neq\mathbf{B}$, then $\mathbf{A}\times\mathbf{B}
\neq\mathbf{B}\times\mathbf{A}$.

\textbf{\textit{Corollary 3:}} If $\mathbf{A}_1, \mathbf{A}_2\in \mathbb{Z}^{n\times n}$ and there exist $\mathbf{A}_1 \times \mathbf{t} \mod (p)=\mathbf{c}_1$ and $\mathbf{A}_2 \times \mathbf{c}_1 \mod (p)=\mathbf{c}_2$, meanwhile, there exist $\mathbf{A}_2 \times \mathbf{t} \mod (p)=\mathbf{c}_3$ and $\mathbf{A}_1 \times \mathbf{c}_3 \mod (p)=\mathbf{c}_4$, then $\mathbf{c}_2\neq\mathbf{c}_4$. 

\textbf{\textit{Proof:}} Please refer to Appendix B.

As shown in \textbf{Algorithm 2}, the newly proposed SUEO algorithm can achieve the purpose of improving the security without increasing the storage overhead of key matrices, just by updating the encryption order of the existing key matrix.

Furthermore, due to the introduction of a large number of matrix multiplication operations in the SUEO algorithm, to optimize the real-time performance of the algorithm, the fast convolutional Winograd \cite{DSWMWinograd} algorithm can be considered to accelerate the protocol design.
Specifically, suppose there are two matrices,
$\mathbf{A}=(\mathbf{a}_1,\mathbf{a}_2,\cdots,
\mathbf{a}_n)^T$ and $\mathbf{B}=(\mathbf{b}_1,\mathbf{b}_2,\cdots,
\mathbf{b}_n)$, where $\mathbf{a}_1=(a_{1},a_{2},\cdots,a_{n})$ and $\mathbf{b}_1=(b_{1},b_{2},\cdots,b_{n})$. Therefore, the elemental value $c_{11}$ for $\mathbf{C}=\mathbf{A}\mathbf{B}$ is as follows
\begin{align}\label{ordinary}
\mathbf{a}_1\bullet\mathbf{b}_1=a_{1}\times b_{1}+a_{2}\times b_{2}+\cdots+a_{n}\times b_{n}.
\end{align}
The calculation of ordinary matrix multiplication is shown in (\ref{ordinary}).
The Winograd algorithm is an effective matrix multiplication method, which is characterized by significantly reducing multiplication operations by adding only a small amount of addition operations. The Winograd algorithm is described as follows:
Let $k=\frac{n}{2}$, when $n$ is an even number,
\begin{align}\label{even}
\mathbf{a}_1\bullet\mathbf{b}_1=&\sum_{i = 1}^{k}\left( a_{2i - 1} +b_{2i} \right)\left( a_{2i} +b_{2i - 1} \right)\nonumber\\
&-\sum_{i = 1}^{k}a_{2i - 1} a_{2i} -\sum_{i = 1}^{k}b_{2i - 1} b_{2i}.
\end{align}
When $n$ is an odd number,
\begin{align}\label{odd}
\mathbf{a}_1\bullet\mathbf{b}_1=&\sum_{i = 1}^{k}\left( a_{2i - 1} +b_{2i} \right)\left( a_{2i} +b_{2i - 1} \right)\nonumber\\
&-\sum_{i = 1}^{k}a_{2i - 1} a_{2i} -\sum_{i = 1}^{k}b_{2i - 1} b_{2i}+a_nb_n.
\end{align}

Comparing (\ref{even}) and (\ref{odd}), it can be found that when $n$ is an odd number, the calculation method is similar to that of an even number, with only one more correction term. For $n$-order square matrix multiplication, the ordinary algorithm contains $n^2(n-1)$ additions and $n^3$ multiplications, while the Winograd algorithm includes $n^2(n+2)$ additions and $n^2(n/2+1)$ multiplications.

\begin{algorithm}[!h]
    \caption{Proposed SUEO encryption algorithm}
    \label{alg:AOA}
        \KwIn{ plaintext $\mathbf{t}$, $N$ initial encryption matrices, modulus $p$, secret value $S$, SUEO index table}
        \KwOut {ciphertext $\mathbf{c}$}
        Initialize set $i=1$ and record the tatal number of elements of the SUEO index table as $Z_{\text{SUEO}}$\;
        Calculate $m_2=\bmod(S,Z_{\text{SUEO}})$\;
        \eIf {$m_2\in [0,N-1]$}{
           one key matrix is needed for encryption\;}
        { 
           set $i=2$\;
           \While{$2\leq i\leq N$}{
             \eIf {$m_2\in \Big[\frac{N*(1-N^{i-1})}{1-N},\frac{N*(1-N^{i})}{1-N}-1\Big]$}
               {$i$ key matrices are needed for encryption\;}
             { 
               $i=i+1$\;}
             }   
        }
\end{algorithm}

\subsection{Proposed DBLTKM encryption algorithm}

\textbf{\textit{Lemma 6:}} If $\mathbf{A}\in \mathbb{Z}^{m\times m}$ and $\mathbf{B}\in \mathbb{Z}^{n\times n}$, then $\det\left(\begin{array}{cc}\mathbf{A}&\\&\mathbf{B}\end{array}\right)
=\det(\mathbf{A})\times \det(\mathbf{B})$.

\textbf{\textit{Lemma 7:}} If $\mathbf{A}\in \mathbb{Z}^{n\times n}$, then $\det(A^T)=\det(A)$.

\textbf{\textit{Corollary 4:}} If $\mathbf{A}\in \mathbb{Z}^{n\times n}$ and there exist $\mathbf{A} \times \mathbf{t} \mod (p)=\mathbf{c}_1$, then $\mathbf{A}^T \times \mathbf{t} \mod (p)=\mathbf{c}_2$.

\textbf{\textit{Proof:}} Please refer to Appendix C.

\textbf{\textit{Corollary 5:}} If $\mathbf{A}_1\in \mathbb{Z}^{m\times m}$ and $\mathbf{A}_2\in \mathbb{Z}^{n\times n}$ and there exist $\mathbf{A}_1 \times \mathbf{t}_1 \mod (p)=\mathbf{c}_1$ and $\mathbf{A}_2 \times \mathbf{t}_2 \mod (p)=\mathbf{c}_2$, then $\left(\begin{array}{cc}\mathbf{A}_1&\\&\mathbf{A}_2\end{array}\right)
\times \left(\begin{array}{c}\mathbf{t}_1\\\mathbf{t}_2\end{array}\right)
\mod (p)=\left(\begin{array}{c}\mathbf{c}_1\\\mathbf{c}_2\end{array}\right)$, where $\mathbf{t}_1\in \mathbb{Z}^{m\times 1}$ and $\mathbf{t}_2\in \mathbb{Z}^{n\times 1}$.

\textbf{\textit{Proof:}} Please refer to Appendix D.

As shown in \textbf{Algorithm 3}, by updating the order of the key matrix on the main diagonal, multiple different key matrices can be formed. It is worth noting that the diversity of this key matrix does not require additional storage of new key matrices, namely the DBLTKM algorithm increases the feasible domain of the key space.

\begin{algorithm}[!h]
    \caption{Proposed DBLTKM encryption algorithm}
    \label{alg:AOA}
        \KwIn{plaintext $\mathbf{t}$, $N$ initial encryption matrices, modulus $p$, secret value $S$, DBLTKM index table}
        \KwOut{ciphertext $\mathbf{c}$} 
        Initialize set $i=1$ and record the total number of elements of the DBLTKM index table as $Z_{\text{DBLTKM}}$\;
        \eIf {plaintext length $l=n$}
           {Calculate $m_3=\bmod(S,Z_{\text{DBLTKM}})$\;
              \If {$m_3\in [0, 2N-1]$}
                {The $m_3$-th key matrix is selected for encryption\;
                Obtain $\mathbf{c}$ by (\ref{keymatrixE})\;
              }
        }{ 
        set $i=2$\;
        }
        \While{$2\leq i\leq N$}{
            \eIf {plaintext length $l=i*n$}
           {Calculate $m_3=\bmod(S,Z_{\text{DBLTKM}})$\;
              \If {$m_3\in \Big[\frac{2N*\big(1-(2N)^{i-1}\big)}{1-2N},\frac{2N*\big(1-(2N)^{i}\big)}{1-2N}-1\Big]$}
                 {The $m_3$-th key matrix is selected for encryption\;
                 Obtain $\mathbf{c}$ by (\ref{keymatrixE})\;
              }
        }
        {$i=i+1$\;
        }
        }
\end{algorithm}

\subsection{Proposed joint AM-SUEO-DBLTKM encryption algorithm}

\begin{figure*}
\centering
	\includegraphics [width=0.99\textwidth]{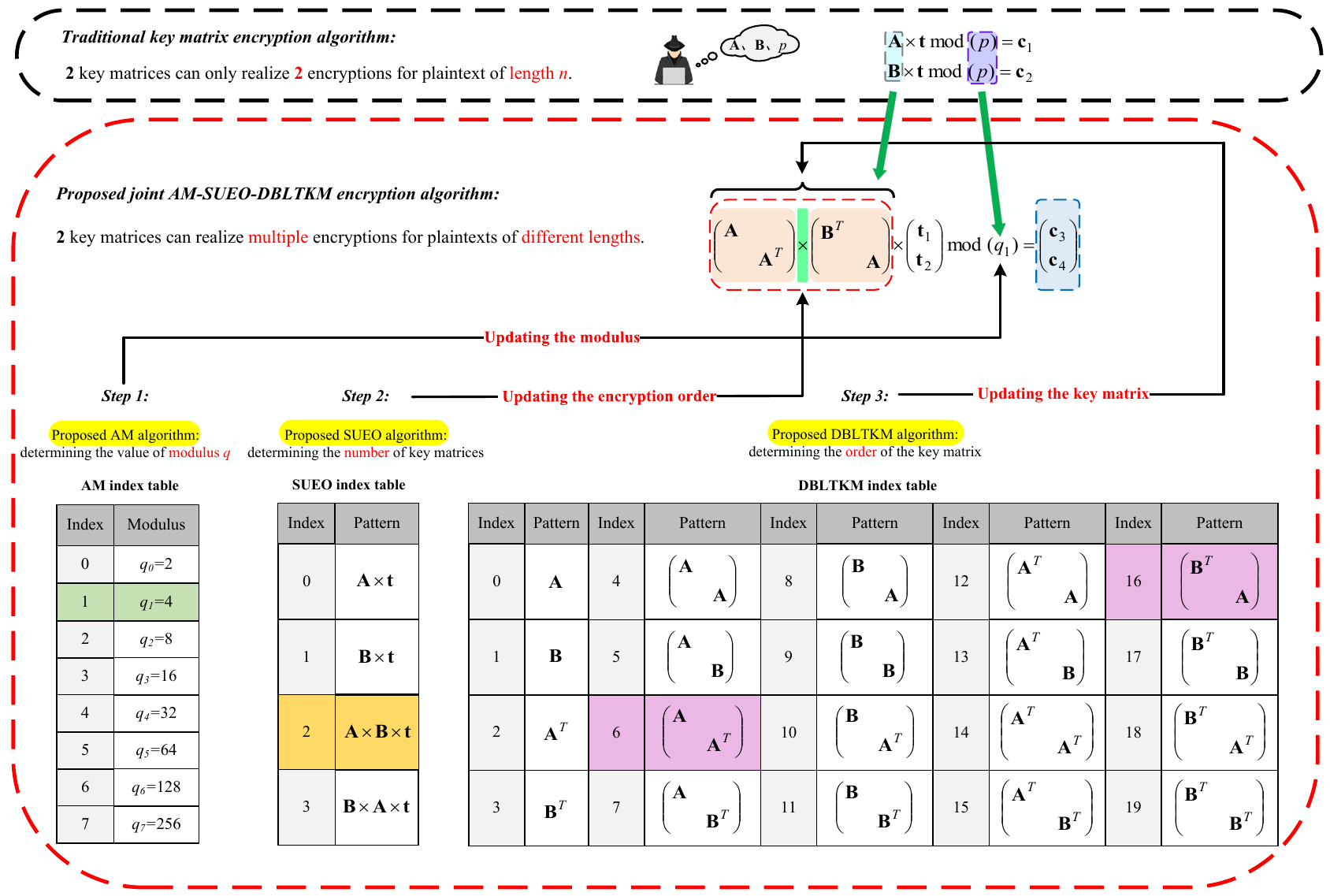}
	\caption{Comparison between the traditional key matrix encryption algorithm and the proposed joint AM-SUEO-DBLTKM encryption algorithm.}
    \label{2key3divisors}
\end{figure*}

The above three algorithms, namely AM algorithm, SUEO algorithm, and DBLTKM algorithm, can be respectively used to implement algorithm encryption and decryption. 
The AM algorithm weakens the correlation between plaintext and ciphertext, the SUEO algorithm improves security, and the DBLTKM algorithm expands the feasible domain of the key space.
To fully leverage the advantages of the three algorithms, a joint AM-SUEO-DBLTKM encryption algorithm is proposed in this subsection.
Taking two key matrices $\mathbf{A}$, $\mathbf{B}$, and modulus $p=16$ as an example, the design principle of the joint AM-SUEO-DBLTKM algorithm is shown in \textbf{Algorithm 4} and Fig.~\ref{2key3divisors}.

\begin{algorithm}[!h]
    \caption{Proposed joint AM-SUEO-DBLTKM encryption algorithm}
    \label{alg:AOA}
        \KwIn{ plaintext $\mathbf{t}$, $N$ initial encryption matrices, modulus $p$, secret value $S$, AM index table, SUEO index table, DBLTKM index table}
        \KwOut {ciphertext $\mathbf{c}$}
        Initialize set $i=1$, and record the total number of elements of the AM index table, the SUEO index table, the DBLTKM index table as $Z_{\text{AM}}$, $Z_{\text{SUEO}}$, $Z_{\text{DBLTKM}}$, respectively\;
        Calculate $m_1=\bmod(S,Z_{\text{AM}})$ and update the modulus $q$ based on AM encryption algorithm\;
        Calculate $m_2=\bmod(S,Z_{\text{SUEO}})$ and update the encryption order based on SUEO encryption algorithm to determine that $N_k$ key matrices are needed for encryption\;
        \While{$1\leq i\leq N_k$}{ 
        Calculate $m_3=\bmod(S,Z_{\text{DBLTKM}})$ and update the key matrix based on DBLTKM encryption algorithm\;
        }
        Obtain $\mathbf{c}$ by (\ref{keymatrixE})\;
         \textbf{return} ciphertext $\mathbf{c}$\;
\end{algorithm}
 
\section{Constructing a two-way AM-SUEO-DBLTKM-RFID authentication protocol}
Based on the joint AM-SUEO-DBLTKM encryption algorithm proposed in the previous section, a two-way RFID security authentication protocol, namely AM-SUEO-DBLTKM-RFID protocol, is constructed in this section.
The initial conditions for the proposed AM-SUEO-DBLTKM-RFID protocol are as follows:

(1) This protocol assumes that communication between the reader and server is wireless and insecure.

\begin{figure*}
\centering
	\includegraphics [width=0.95\textwidth]{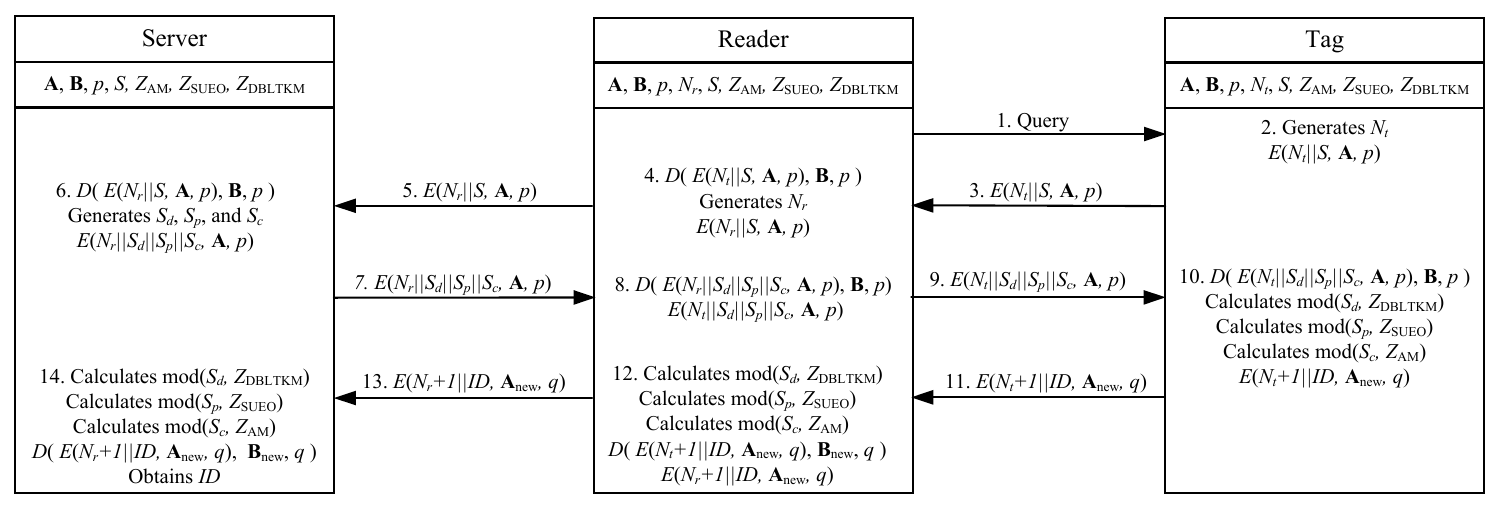}
	\caption{Proposed two-way AM-SUEO-DBLTKM-RFID authentication protocol.}
    \label{DBKM-SUEO-SUM-RFID}
\end{figure*}

(2) To initiate the protocol and carry out subsequent procedures, variables that need to be pre-stored in the server, reader, and tag are shown in Fig.~\ref{DBKM-SUEO-SUM-RFID}.
The symbols involved in the proposed protocol are indicated in Table \ref{SymbolNotations}. 
According to \cite{FanPermutation}
\cite{ShihabLightweight}, the RNs, secret values, modulus, and IDs involved in the protocol authentication process are assumed to be 128 bits, and each element of the key matrix is 64 bits.
Taking two key matrices and modulus $p=16$ as an example, the AM index table, SUEO index table, and DBLTKM index table are described in Fig.~\ref{2key3divisors}.

\begin{table}[h]
    \centering
    \footnotesize
    \renewcommand{\arraystretch}{1.0}
    \setlength{\tabcolsep}{0.5pt}
    \caption{Notations used in the protocol's description}
    \label{SymbolNotations}
    \scalebox{0.85}{
    \begin{tabular}{cc}
    \toprule
      \textbf{Notations}  &\textbf{Meaning}  \\ \midrule
      $N_t$ & The RN which is generated by tag\\
      $N_r$ & The RN which is generated by reader\\
      $S$ & Secret value\\
      $S_d$ & The secret value used to determine the construction of block key matrix\\
      $S_p$ & The secret value used to determine the encryption order\\
      $S_c$ & The secret value used to determine the selection of modulus \\
      $p$ & Initial modulus \\
      $q$ & Updated modulus\\
      $N$ & The total number of key matrices \\
      $Z_\text{DBLTKM}$ & The total number of elements of the DBLTKM index table \\
     $Z_\text{SUEO}$ & The total number of elements of the SUEO index table\\
      $Z_\text{AM}$ & The total number of elements of the AM index table\\
      $\mathbf{A}$, $\mathbf{A}_{\text{new}}$ & Initial encryption matrix and updated encryption matrix  \\
      $\mathbf{B}$, $\mathbf{B}_{\text{new}}$ & Initial decryption matrix and updated decryption matrix\\
    \bottomrule
    \end{tabular}
    }
\end{table}

The specific authentication details of the AM-SUEO-DBLTKM-RFID protocol are as follows:

1. The reader sends a ``Query'' to the tag.

2. The tag responds to the reader and uses its internal pseudo RN generator to generate $N_t$. Subsequently, the tag uses the encryption matrix $\mathbf{A}$ and modulus $p$ to encrypt $N_t\|S$, denoted as $E(N_t\|S, \mathbf{A}, p)$.

3. The tag sends $E(N_t\|S, \mathbf{A}, p)$ to the reader.

4. The reader decrypts $E(N_t\|S, \mathbf{A}, p)$ sent by the tag and obtains the secret value $S$. If $S$ can be queried, it indicates that the reader has successfully authenticated the tag and accepted $N_t$. Otherwise, the protocol is terminated. Then, the reader generates $N_r$ and uses $\mathbf{A}$ and $p$ to encrypt $N_r\|S$, denoted as $E(N_r\|S, \mathbf{A}, p)$.

5. The reader sends $E(N_r\|S, \mathbf{A}, p)$ to the server.

6. The server decrypts $E(N_r\|S, \mathbf{A}, p)$ sent by the reader and obtains the secret value $S$. If $S$ can be queried, it indicates that the server has successfully authenticated the reader and accepted $N_r$. Otherwise, the protocol is terminated. Subsequently, the new secret values $S_d$, $S_p$, and $S_c$ are generated, the server uses $\mathbf{A}$ and $p$ to encrypt $N_r\|S_d\|S_p\|S_c$, denoted as $E(N_r\|S_d\|S_p\|S_c, \mathbf{A}, p)$.

7. The server sends $E(N_r\|S_d\|S_p\|S_c, \mathbf{A}, p)$ to the reader.

8. The reader decrypts $E(N_r\|S_d\|S_p\|S_c, \mathbf{A}, p)$ sent by the server and obtains $N_r$. If $N_r$ is equal to the previous one, which indicates that the reader has successfully authenticated the server. Correspondingly, the new secret values $S_d$, $S_p$, and $S_c$ are accepted. Otherwise, the protocol is terminated. Then, the reader uses $\mathbf{A}$ and $p$ to encrypt $N_t\|S_d\|S_p\|S_c$, denoted as $E(N_t\|S_d\|S_p\|S_c, \mathbf{A}, p)$.

9. The reader sends $E(N_t\|S_d\|S_p\|S_c, \mathbf{A}, p)$ to the tag.

10. The tag decrypts $E(N_t\|S_d\|S_p\|S_c, \mathbf{A}, p)$ sent by the reader and obtains $N_t$. If $N_t$ is equal to the previous one, which indicates that the tag has successfully authenticated the reader. Correspondingly, the new secret values $S_d$, $S_p$, and $S_c$ are accepted. Otherwise, the protocol is terminated. Subsequently, the value of $\operatorname{mod}(S_d, Z_{\text{DBLTKM}})$ is calculated to determine the construction method of diagonal block key matrix. Similarly, the value of $\operatorname{mod}(S_p, Z_{\text{SUEO}})$ is calculated to determine the encryption order, and the value of $\operatorname{mod}(S_c, Z_{\text{AM}})$ is calculated to determine the selection of modulus. When the new key matrix $\mathbf{A}_{\text{new}}$, new modulus $q$, and encryption order are obtained, the tag uses $\mathbf{A}_{\text{new}}$ and $q$ to encrypt $N_t+1\|ID$, denoted as $E(N_t+1\|ID, \mathbf{A}_{\text{new}}, q)$.

11. The tag sends $E(N_t+1\|ID, \mathbf{A}_{\text{new}}, q)$ to the reader.

12. The values of $\operatorname{mod}(S_d, Z_{\text{DBLTKM}})$, $\operatorname{mod}(S_p, Z_{\text{SUEO}})$, and $\operatorname{mod}(S_c, Z_{\text{AM}})$ are calculated by the reader. Then, the reader decrypts $E(N_t+1\|ID, \mathbf{A}_{\text{new}}, q)$ based on the newly obtained $\mathbf{B}_{\text{new}}$ and $q$. If $N_t$ is equal to the previous one, $ID$ is obtained. The reader uses $\mathbf{A}_{\text{new}}$ and $q$ to encrypt $N_r+1\|ID$, denoted as $E(N_r+1\|ID, \mathbf{A}_{\text{new}}, q)$.

13. The reader sends $E(N_r+1\|ID, \mathbf{A}_{\text{new}}, q)$ to the server.

14. The values of $\operatorname{mod}(S_d, Z_{\text{DBLTKM}})$, $\operatorname{mod}(S_p, Z_{\text{SUEO}})$, and $\operatorname{mod}(S_c, Z_{\text{AM}})$ are calculated by the server. Then, the server decrypts $E(N_r+1\|ID, \mathbf{A}_{\text{new}}, q)$ based on the newly obtained $\mathbf{B}_{\text{new}}$ and $q$. If $N_r$ is equal to the previous one, $ID$ is obtained.

\section{Performance Evaluation}

In this section, to start with, the BAN logic of the proposed AM-SUEO-DBLTM-RFID protocol is demonstrated in subsection A.
Subsequently, in subsection B, the security analysis of the AM-SUEO-DBLTM-RFID protocol is completed.
In addition, the tag storage overhead computation is presented in subsection C.
Finally, the tag computational complexity is presented in subsection D.

\subsection{BAN logic}
BAN logic \cite{SABAN} is one species of modal logic. 
It has been used to verify the security of numerous authentication protocols.
We formally analyze the proposed AM-SUEO-DBLTM-RFID protocol through BAN logic. 

The syntax and semantics of BAN logic involved in this protocol are presented in Table \ref{BANnotations}.
\begin{table}[h]
    \centering
    \footnotesize
    \renewcommand{\arraystretch}{1.0}
    \setlength{\tabcolsep}{8pt}
    \caption{BAN logic notations}
    \label{BANnotations}
    \scalebox{1}{
    \begin{tabular}{cc}
    \toprule
      \textbf{Notations}  &\textbf{Meaning}  \\ \midrule
      $P \mid \equiv X$ & $P$ believes $X$\\
$P \vartriangleleft  X$ & $P$ receives $X$\\
$P \mid \sim X$ & $P$ sends $X$\\
$P \mid \Rightarrow X$ & $P$ has jurisdiction over $X$\\
$\#(X)$ & $X$ is fresh\\
$\{X\}_k$ & $X$ is encrypted by the secret $k$\\
$P \stackrel{k}{\longleftrightarrow} Q$ & $P$ and $Q$ have a shared secret $k$\\
    \bottomrule
    \end{tabular}
    }
\end{table}

Some reasoning rules of BAN logic involved in this protocol are described in Table \ref{R1-R4}.
\begin{table}[h]
    \centering
    \footnotesize
    \renewcommand{\arraystretch}{1.2}
    \setlength{\tabcolsep}{8pt}
    \caption{BAN logic rules}
    \label{R1-R4}
    \scalebox{1}{
    \begin{tabular}{cc}
    \toprule
      \textbf{Notations}  &\textbf{Meaning}  \\ \midrule
     R1 (Message-meaning rule) & $\frac{P\mid \equiv Q \stackrel{k}{\longleftrightarrow} P,P \vartriangleleft \{X\}_k}{P \mid \equiv Q \mid \sim X}$\\
R2 (Nonce-verification rule) & $\frac{P \mid \equiv \#(X),P \mid \equiv Q \mid \sim X}{P \mid \equiv Q \mid \equiv X}$\\
R3 (Jurisdiction rule) & $\frac{P \mid \equiv Q \mid \Rightarrow X, P \mid \equiv Q \mid \equiv X}{P \mid \equiv X}$\\
R4 (Fresh rule) & $\frac{P \mid \equiv \#(X)}{P \mid \equiv \#(X,Y)}$\\
    \bottomrule
    \end{tabular}
    }
\end{table}

The proposed authentication scheme includes message exchange between three entities, where $S$ denotes the server, $R$ represents the reader, and $T$ is the tag. The idealized descriptions of the proposed protocol are shown Table \ref{TRS}. 
\begin{table}[h]
    \centering
    \footnotesize
    \renewcommand{\arraystretch}{1.0}
    \setlength{\tabcolsep}{8pt}
    \caption{An idealized description of the protocol}
    \label{TRS}
    \scalebox{1}{
    \begin{tabular}{cc}
    \toprule
      \textbf{Notations}  &\textbf{Meaning}  \\ \midrule
      $T\rightarrow R$ & \ $\{N_t\|S\}_{\mathbf{A}, p}$\\
      $R\rightarrow S$ & \ $\{N_r\|S\}_{\mathbf{A}, p}$\\

$S\rightarrow R$ & \ $\{N_r\|S_d\|S_p\|S_c\}_{\mathbf{A}, p}$\\

$R\rightarrow T$ & \ $\{N_t\|S_d\|S_p\|S_c\}_{\mathbf{A}, p}$\\

$T\rightarrow R$ & \ $\{N_t+1\|ID\}_{\mathbf{A}_{\text{new}}, q}$\\

$R\rightarrow S$ & \ $\{N_r+1\|ID\}_{\mathbf{A}_{\text{new}}, q}$\\
    \bottomrule
    \end{tabular}
    }
\end{table}

According to Table \ref{TRS}, the model of idealized messages for proposed protocol are displayed in Table \ref{M1-M6}.
\begin{table}[h]
    \centering
    \footnotesize
    \renewcommand{\arraystretch}{1.0}
    \setlength{\tabcolsep}{8pt}
    \caption{Model of idealized messages for proposed protocol}
    \label{M1-M6}
    \scalebox{1}{
    \begin{tabular}{cc}
    \toprule
      \textbf{Notations}  &\textbf{Meaning}  \\ \midrule
      M1 & \ $ R \vartriangleleft \ \{N_t\|S\}_{\mathbf{A}, p}$\\

M2 & \ $ S \vartriangleleft \ \{N_r\|S\}_{\mathbf{A}, p}$\\

M3 & \ $ R \vartriangleleft \ \{N_r\|S_d\|S_p\|S_c\}_{\mathbf{A}, p}$\\

M4 & \ $ T \vartriangleleft \ \{N_t\|S_d\|S_p\|S_c\}_{\mathbf{A}, p}$\\

M5 & \ $ R \vartriangleleft \ \{N_t+1\|ID\}_{\mathbf{A}_{\text{new}}, q}$\\

M6 & \ $ S \vartriangleleft \ \{N_r+1\|ID\}_{\mathbf{A}_{\text{new}}, q}$\\
    \bottomrule
    \end{tabular}
    }
\end{table}

The proposed protocol’s initialization assumptions for BAN logic proof are listed in Table \ref{A1-A16}.
\begin{table}[h]
    \centering
    \footnotesize
    \renewcommand{\arraystretch}{1.0}
    \setlength{\tabcolsep}{8pt}
    \caption{The initialization assumption}
    \label{A1-A16}
    \scalebox{1}{
    \begin{tabular}{cc}
    \toprule
      \textbf{Notations}  &\textbf{Meaning}  \\ \midrule
     A1 & \ $R \mid \equiv T \stackrel{key}{\longleftrightarrow} R$\\

A2 & \ $R \mid \equiv \# (N_t)$\\

A3 & \ $R \mid \equiv T \mid \Rightarrow \{N_t\|S\}$\\

A4 & \ $S \mid \equiv R \stackrel{key}{\longleftrightarrow} S$\\

A5 & \ $S \mid \equiv \# (N_r)$\\

A6 & \ $S \mid \equiv R \mid \Rightarrow \{N_r\|S\}$\\

A7 & \ $R \mid \equiv S \stackrel{key}{\longleftrightarrow} R$\\

A8 & \ $R \mid \equiv \# (S_d\|S_p\|S_c)$\\

A9 & \ $R \mid \equiv S \mid \Rightarrow \{N_r\|S_d\|S_p\|S_c\}$\\

A10 & \ $T \mid \equiv R \stackrel{key}{\longleftrightarrow} T$\\

A11 & \ $T \mid \equiv \# (S_d\|S_p\|S_c)$\\

A12 & \ $T \mid \equiv R \mid \Rightarrow \{N_t\|S_d\|S_p\|S_c\}$\\

A13 & \ $R \mid \equiv \# (ID)$\\

A14 & \ $R \mid \equiv T \mid \Rightarrow \{N_t+1\|ID\}$\\

A15 & \ $S \mid \equiv \# (ID)$\\

A16 & \ $S \mid \equiv R \mid \Rightarrow \{N_r+1\|ID\}$\\
    \bottomrule
    \end{tabular}
    }
\end{table}

The inference goals of the proposed protocol are depicted in Table \ref{G1-G6}.
\begin{table}[h]
    \centering
    \footnotesize
    \renewcommand{\arraystretch}{1.0}
    \setlength{\tabcolsep}{8pt}
    \caption{BAN logic inference goal}
    \label{G1-G6}
    \scalebox{1}{
    \begin{tabular}{cc}
    \toprule
      \textbf{Notations}  &\textbf{Meaning}  \\ \midrule
     G1 & \ $R \mid \equiv \{N_t\|S\}$\\

G2 & \ $S \mid \equiv \{N_r\|S\}$\\

G3 & \ $R \mid \equiv \{N_r\|S_d\|S_p\|S_c\}$\\
G4 & \ $T \mid \equiv \{N_t\|S_d\|S_p\|S_c\}$\\
G5 & \ $R \mid \equiv \{N_t+1\|ID\}$\\
G6 & \ $S \mid \equiv \{N_r+1\|ID\}$\\
    \bottomrule
    \end{tabular}
    }
\end{table}

The specific reasoning process of the proposed protocol is as follows:

From M1, A1, and R1, we derive
\begin{equation}\label{1}
R \mid \equiv T \mid \sim \{N_t\|S\}.
\end{equation}

From A2 and R4, we derive
\begin{equation}\label{2}
 R \mid \equiv \# \{N_t\|S\}.
\end{equation}

From (\ref{1}), (\ref{2}), and R2, we derive
\begin{equation}\label{3}
 R \mid \equiv T \mid \equiv \{N_t\|S\}.
\end{equation}

From (\ref{3}), A3, and R3, we derive
\begin{equation}\label{4}
 R \mid \equiv \{N_t\|S\}.
\end{equation}

(\ref{4}) shows that the inference goal in G1.

From M2, A4, and R1, we derive
\begin{equation}\label{5}
S \mid \equiv R \mid \sim \{N_r\|S\}.
\end{equation}

From A5 and R4, we derive
\begin{equation}\label{6}
 S \mid \equiv \# \{N_r\|S\}.
\end{equation}

From (\ref{5}), (\ref{6}), and R2, we derive
\begin{equation}\label{7}
 S \mid \equiv R \mid \equiv \{N_r\|S\}.
\end{equation}

From (\ref{7}), A6, and R3, we derive
\begin{equation}\label{8}
 S \mid \equiv \{N_r\|S\}.
\end{equation}

(\ref{8}) shows that the inference goal in G2.

From M3, A7, and R1, we derive
\begin{equation}\label{9}
 R \mid \equiv S \mid \sim \{N_r\|S_d\|S_p\|S_c\}.
\end{equation}

From A8 and R4, we derive
\begin{equation}\label{10}
R \mid \equiv \# (N_r\|S_d\|S_p\|S_c).
\end{equation}

From (\ref{9}), (\ref{10}), and R2, we derive
\begin{equation}\label{12}
R \mid \equiv S \mid \equiv \{N_r\|S_d\|S_p\|S_c\}.
\end{equation}

From A9, (\ref{12}), and R3, we derive
\begin{equation}\label{13}
R \mid \equiv \{N_r\|S_d\|S_p\|S_c\}.
\end{equation}

(\ref{13}) shows that the inference goal in G3.

From M4, A10, and R1, we derive
\begin{equation}\label{13-1}
T \mid \equiv R \mid \sim \{N_t\|S_d\|S_p\|S_c\}.
\end{equation}

From A11 and R4, we derive
\begin{equation}\label{14}
 T \mid \equiv \# \{N_t\|S_d\|S_p\|S_c\}.
\end{equation}

From (\ref{13-1}), (\ref{14}), and R2, we derive
\begin{equation}\label{15}
 T \mid \equiv R \mid \equiv \{N_t\|S_d\|S_p\|S_c\}.
\end{equation}

From (\ref{15}), A12, and R3, we derive
\begin{equation}\label{16}
 T \mid \equiv \{N_t\|S_d\|S_p\|S_c\}.
\end{equation}

(\ref{16}) shows that the inference goal in G4.

From M5, A1, and R1, we derive
\begin{equation}\label{17}
R \mid \equiv T \mid \sim \{N_t+1\|ID\}.
\end{equation}

From A13 and R4, we derive
\begin{equation}\label{18}
 R \mid \equiv \# \{N_t+1\|ID\}.
\end{equation}

From (\ref{17}), (\ref{18}), and R2, we derive
\begin{equation}\label{19}
 R \mid \equiv T \mid \equiv \{N_t+1\|ID\}.
\end{equation}

From (\ref{19}), A14, and R3, we derive
\begin{equation}\label{20}
 R \mid \equiv \{N_t+1\|ID\}.
\end{equation}

(\ref{20}) shows that the inference goal in G5.

From M6, A4, and R1, we derive
\begin{equation}\label{21}
S \mid \equiv R \mid \sim \{N_r+1\|ID\}.
\end{equation}

From A15 and R4, we derive
\begin{equation}\label{22}
 S \mid \equiv \# \{N_r+1\|ID\}.
\end{equation}

From (\ref{21}), (\ref{22}), and R2, we derive
\begin{equation}\label{23}
 S \mid \equiv R \mid \equiv \{N_r+1\|ID\}.
\end{equation}

From (\ref{23}), A16, and R3, we derive
\begin{equation}\label{24}
 S \mid \equiv \{N_r+1\|ID\}.
\end{equation}

(\ref{24}) shows that the inference goal in G6.

\subsection{Security Analysis}

The security comparison between the newly proposed AM-SUEO-DBLTM-RFID protocol and several other protocols is shown in Table \ref{Protocol-security}.
\begin{table}[h]
    \centering
    \renewcommand{\arraystretch}{1.0}
    \setlength{\tabcolsep}{9pt}
    \caption{Protocol security comparison}
    \label{Protocol-security}
    \scalebox{0.7}{
    \begin{tabular}{ccccc}
    \toprule
      \textbf{Protocol}  &\textbf{\cite{FanPermutation}}
      &\textbf{\cite{YLuo}}&\textbf{\cite{WANGEfficient}}&\textbf{Our}  \\ \midrule
     Mutual authentication  & YES & YES & NO& YES\\
  Location tracking  & YES & YES & YES & YES\\
  DoS  & NO & YES & YES & YES\\
  Impersonation attack & NO & YES & YES& YES \\
  Man-in-the-middle attack  & NO & NO & YES & YES\\
  Replay attack & YES & YES & YES & YES\\
  De-synchronization  & YES & YES & YES & YES\\
  Forward secrecy  & YES & YES & YES & YES\\
    \bottomrule
    \end{tabular}
    }
\end{table}

(1) Mutual authentication: In the process of protocol authentication, mutual authentication between servers, readers, and tags is achieved through secret values and RNs. Among them, the reader authentication tag and the server authentication reader are verified using the shared secret value $S$.
In addition, the reader authentication server, as well as the tag authentication reader, are all confirmed through their own generated RNs.
The protocol will only continue if both parties involved in the information exchange have successfully verified it. Otherwise, the protocol will terminate immediately.
Therefore, this protocol achieves mutual authentication among the server, reader, and tag.

(2) Location tracking: Firstly, the data sent during the authentication process of the AM-SUEO-DBLTM-RFID protocol contains RNs and secret values. Secondly, the key matrix is determined by the secret values. Moreover, the modulus, encryption order and key matrix during the encryption and decryption process are both self updating. Therefore, the feedback information between tags and readers is random, and attackers are unable to locate and track tags.

(3) DoS attack \cite{ZhangIoT2023a}: If an attacker sends a large amount of false or incorrect information, causing the system to malfunction or interrupting normal communication, it can lead to a DoS attack. 
However, in our proposed protocol, the reader needs to query whether the secret value $S$ sent by the tag is consistent with the secret value stored by the reader itself, and then decide whether to carry out the next step of communication. This ``query before authentication'' method effectively resists DoS attacks.

(4) Impersonation attack: First of all, when the attacker attempts to disguise as a tag, the reader cannot find the corresponding secret value $S$ when querying the backend database, and then the attacker is marked as an illegal tag. Secondly, when the attacker disguises himself as a reader, the tag cannot successfully verify the RN $N_t$, so the attacker is marked as an illegal reader. Similarly, the reader and server can also defend against impersonation attacks.

(5) Man-in-the-middle attack: Before an attacker can implement a man-in-the-middle attack, they need to know the RNs, secret values, and even key matrices sent between the reader and the tag. However, this protocol updates the secret values and key matrix before each authentication, thus avoiding man-in-the-middle attacks.

\begin{figure}[h]
\centering
 	\includegraphics [width=0.6\textwidth]{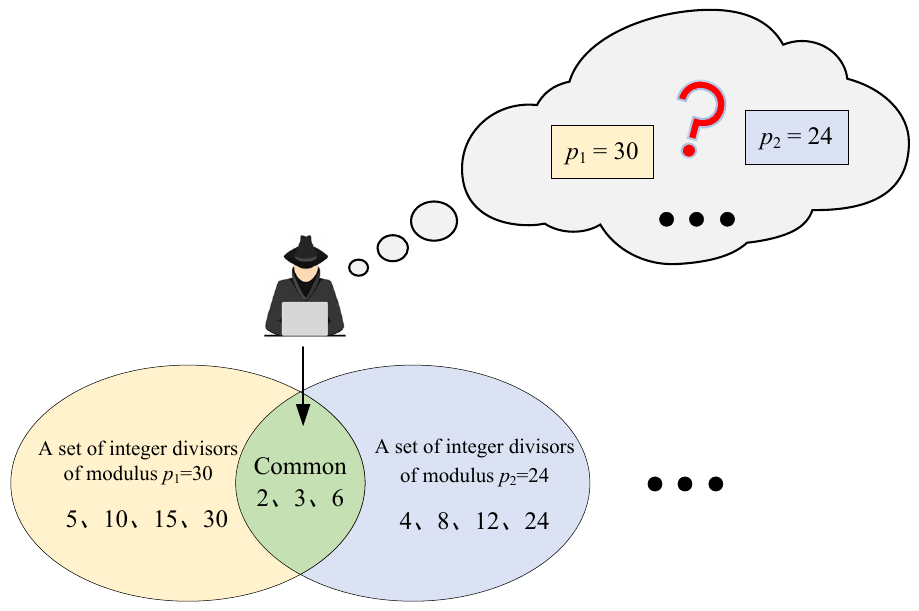}
	\caption{Attackers cannot infer $p$ through $q$.}
    \label{Pbianmo}
\end{figure}
(6) Replay attack: The authentication data between tags and readers remains fresh due to the presence of RNs. Moreover, the update of the key matrix updates with the update of the secret value, so the attacker cannot derive the current value from the information intercepted in the previous round. 
As shown in Fig.~\ref{Pbianmo}, even if the attacker steals the currently used modulus $q$, the attacker cannot infer the last used modulus $p$. Because the same $q$ may be an integer divisor of multiple modulus $p$. Therefore, the attacker is unable to perform replay attacks.

(7) De-synchronization: An Attacker would need to carry out de-synchronization attacks by blocking or modifying a portion or all of the data in communication between tags and readers, causing them to become out of sync \cite{LA}. However, the proposed protocol communicates only after successful authentication. Moreover, both the reader and tag pre-store the AM index table, SUEO index table, and DBLTKM index table, ensuring that even if the current authentication fails, the previous round of key matrix can be used for recalculation, which further defends de-synchronization attacks.

(8) Forward secrecy: The data transmitted by this protocol contains RNs and secret values, and both the modulus and key matrix can achieve self updating. Therefore, as shown in Fig.~\ref{Pbianmo}, attackers cannot derive the previous authentication data from this currently message, indicating that the protocol has forward security.

\subsection{Tag Storage Overhead}
In general, it is believed that the storage space of servers and readers is much stronger than that of tags. Therefore, in this subsection, we will focus on the analysis of the savings of the tag storage space of the proposed 7 key matrix encryption algorithms.

In the traditional encryption and decryption algorithm based on key matrix, $N$ number of $n$-order key matrices can complete a total of $N$ encryptions and decryptions of the plaintext length $n$. However, for the newly proposed AM algorithm, the traditional constant modulus is extended to variable modulus, which can realize $N\times Z_{\text{AM}}$ encryptions and decryptions. In this case, the number of variables that need to be pre-stored by the tag is:
\begin{align}
S_{\text{AM}}=N\times n^2+Z_{\text{AM}}.
\end{align}
If there is no AM algorithm, traditional key matrix encryption algorithms need to store multiple sets of encryption and decryption matrices in advance in order to achieve the same $N\times Z_{\text{AM}}$ encryptions and decryptions. In this case, the storage space required to occupy the tag is:
\begin{align}
S_{\text{wihout AM}}=N\times Z_{\text{AM}}\times n^2+1.
\end{align}

As shown in Fig.~\ref{2key3divisors}, in the newly proposed SUEO algorithm, $N$ number of $n$-order key matrices can accomplish a total of $Z_{\text{SUEO}}$ encryptions and decryptions of plaintexts of length $n$, where
\begin{align}
Z_{\text{SUEO}}=C_N^1\times 1!+C_N^2\times 2!+\cdots+C_N^N\times N!.
\end{align}
In this case, the number of variables that need to be pre-stored by the tag is
\begin{align}
&S_{\text{SUEO}}=N\times n^2+Z_{\text{SUEO}}+1\\
&=N\times n^2+C_N^1\times 1!+C_N^2\times 2!+\cdots+C_N^N\times N!+1.\nonumber
\end{align}
In traditional key matrix encryption algorithms, to achieve the same $Z_{\text{SUEO}}$ encryptions, the number of variables stored in the tag is as follows
\begin{align}
&S_{\text{without SUEO}}=Z_{\text{SUEO}}\times n^2+1\\
&=\bigg(C_N^1\times 1!+C_N^2\times 2!+\cdots+C_N^N\times N!\bigg)\times n^2+1.\nonumber
\end{align}

In addition, as depicted in Fig.~\ref{2key3divisors}, when the proposed DBLTKM algorithm is employed, $N$ number $n$-order key matrices can realize $Z_{\text{DBLTKM}}$ encryptions, where
\begin{align}
Z_{\text{DBLTKM}}&=(2N)+(2N)^2+\cdots+(2N)^N\nonumber\\
&=\frac{2N\times\big(1-(2N)^N\big)}{1-2N}.
\end{align}
In this scenario, the number of variables that need to be pre-stored by the tag is as follows
\begin{align}
S_{\text{DBLTKM}}
&=N\times n^2+Z_{\text{DBLTKM}}+1\nonumber\\
&=N\times n^2+\frac{2N\times\big(1-(2N)^N\big)}{1-2N}+1.
\end{align}
In contrast, without the DBLTKM algorithm, traditional key matrix encryption algorithms require the number of tag storage variables to be
\begin{align}
S_{\text{without DBLTKM}}
=&(2N)\times (n)^2+(2N)^2\times (2n)^2+\cdots\nonumber\\
&+(2N)^N\times (N n)^2+1.
\end{align}

To sum up, the tag storage space savings of the proposed AM algorithm, SUEO algorithm, and DBLTKM algorithm are analyzed. Next, the tag storage overhead of the mixtures of the three algorithms are computed.

Firstly, for the AM-SUEO algorithm, it needs tags to pre-store AM index table and SUEO index table, at which time the storage overhead of the tag is as follows:
\begin{align}
S_{\text{AM-SUEO}}=&N\times n^2+Z_{\text{AM}}+Z_{\text{SUEO}}\nonumber\\
=&N\times n^2+Z_{\text{AM}}+C_N^1\times 1!+C_N^2\times 2!+\cdots\nonumber\\
&+C_N^N\times N!.
\end{align}
If there is no AM-SUEO algorithm, in order to complete the same $Z_{\text{AM}}\times Z_{\text{SUEO}}$ encryptions, the number of variables stored in the tag is:
\begin{align}
&S_{\text{without AM-SUEO}}=Z_{\text{AM}}\times Z_{\text{SUEO}}\times n^2+1\\
&=Z_{\text{AM}}\times\bigg(C_N^1\times 1!+C_N^2\times 2!+\cdots
+C_N^N\times N!\bigg)
\times n^2+1.\nonumber
\end{align}

Secondly, similar to the calculation above, for using the AM-DBLTKM algorithm and not using the AM-DBLTKM algorithm, the total number of variables to be stored in the tag is 
\begin{align}
S_{\text{AM-DBLTKM}}&=N\times n^2+Z_{\text{AM}}+Z_{\text{DBLTKM}}\nonumber\\
&=N\times n^2+Z_{\text{AM}}+\frac{2N\times\big(1-(2N)^N\big)}{1-2N},
\end{align}
and
\begin{align}
&S_{\text{without AM-DBLTKM}}=Z_{\text{AM}}\times \bigg\{(2N)\times (n)^2\nonumber\\
&+(2N)^2\times (2n)^2+\cdots+(2N)^N\times (N n)^2\bigg\}+1,
\end{align}
respectively.

Thirdly, when SUEO-DBLTKM algorithm is adopted, the storage space of the tags is also saved, which is
\begin{align}
S_{\text{SUEO-DBLTKM}}&=N \times n^2+Z_{\text{SUEO}}+Z_{\text{DBLTKM}}\nonumber\\
&=N \times n^2+C_N^1\times 1!+C_N^2\times 2!+\cdots\nonumber\\
&+C_N^N\times N!+\frac{2N\times\big(1-(2N)^N\big)}{1-2N}.
\end{align}
Correspondingly, when there is no SUEO-DBLTKM algorithm, the tag needs to store the following number of variables
\begin{align}
&S_{\text{without SUEO-DBLTKM}}=
\bigg(C_{2N}^1+C_{2N}^2+\cdots
+C_{2N}^N\bigg)\times n^2\nonumber\\
&+\bigg(C_{(2N)^2}^1+C_{(2N)^2}^2+\cdots
+C_{(2N)^2}^N\bigg)\times (2n)^2+\cdots
\nonumber\\
&+\bigg(C_{(2N)^N}^1+C_{(2N)^N}^2+\cdots
+C_{(2N)^N}^N\bigg)\times (Nn)^2.
\end{align}

For the proposed AM-SUEO-DBLTKM algorithm, it requires the number of tag storage variables to be
\begin{align}
S_{\text{AM-SUEO-DBLTKM}}&= N\times n^2+Z_{\text{AM}}+Z_{\text{SUEO}}
+Z_{\text{DBLTKM}}\nonumber\\
&=N\times n^2+Z_{\text{AM}}+C_N^1\times 1!+C_N^2\times 2!\nonumber\\
&+\cdots+C_N^N\times N!+
\frac{2N\times\big(1-(2N)^N\big)}{1-2N}.
\end{align}
However, the corresponding traditional key matrix algorithm for this scenario requires the number of variables stored in the tag as follows
\begin{align}
&S_{\text{without AM-SUEO-DBLTKM}}=
\bigg(C_{2N}^1+C_{2N}^2+\cdots
+C_{2N}^N\bigg)\nonumber\\
&\times Z_{\text{AM}} \times n^2
+\bigg(C_{(2N)^2}^1+C_{(2N)^2}^2+\cdots
+C_{(2N)^2}^N\bigg)
\times Z_{\text{AM}}\nonumber\\
&\times (2n)^2+\cdots
+\bigg(C_{(2N)^N}^1+C_{(2N)^N}^2+\cdots
+C_{(2N)^N}^N\bigg)\nonumber\\
&\times Z_{\text{AM}}\times (Nn)^2+1.
\end{align}

Let's define $K_{\text{AM}}$ as the tag storage saving ratio of the AM algorithm, where
\begin{align}
K_{\text{AM}}=\frac{(S_{\text{without AM}}-S_{\text{AM}})}{S_{\text{without AM}}}\times 100\%.
\end{align}
Similarly, we can obtain $K_{\text{SUEO}}$, $K_{\text{DBLTKM}}$, $K_{\text{AM-SUEO}}$, $K_{\text{AM-DBLTKM}}$, $K_{\text{SUEO-DBLTKM}}$, and $K_{\text{AM-SUEO-DBLTKM}}$.

\section{Numerical results and discussions}
In what follows, we will present numerical simulations to evaluate the performance of the proposed seven encryption algorithms. In general, the total number of variables pre-stored in tags is related to the values of $Q$, $N$, and $n$.
\begin{figure}[h]
\centering
	\includegraphics [width=0.9\textwidth]{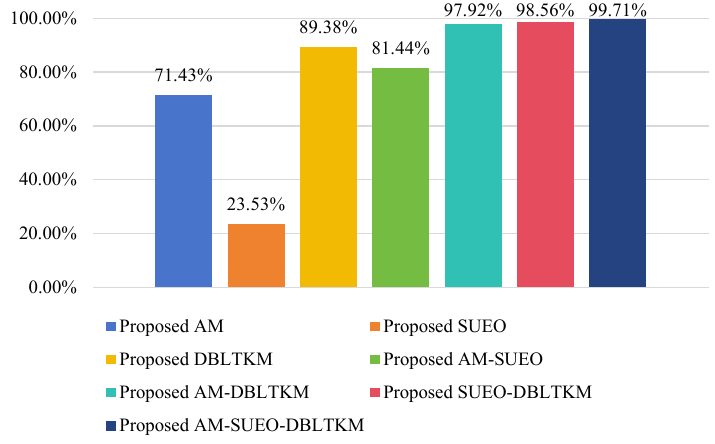}
	\caption{Tag storage saving ratio $K$ when the total number of variable modulus $Q=3$, the number of key matrices $N=2$ and the plaintext length $n=2$.}
    \label{zhuzhuangtu}
\end{figure}

Fig.~\ref{zhuzhuangtu} plots the histogram of the tag storage space savings of the proposed three encryption algorithms, namely AM, SUEO, DBLTKM, and their mixtures where $Q=3$, and $N=n=2$. Clearly, they achieve different savings on the tag storage space. Observing this figure, using only the proposed single SUEO algorithm can save at least 23.53\% of tag storage space, while the proposed joint AM-SUEO-DBLTKM algorithm can save up to 99.71\% of tag storage space. 
However, not all mixture methods make a significant performance enhancement over single ones. For example, the proposed DBLTKM algorithm can achieve better performance improvement than the joint AM-SUEO algorithm.

\begin{figure}[h]
\centering
	\includegraphics [width=0.9\textwidth]{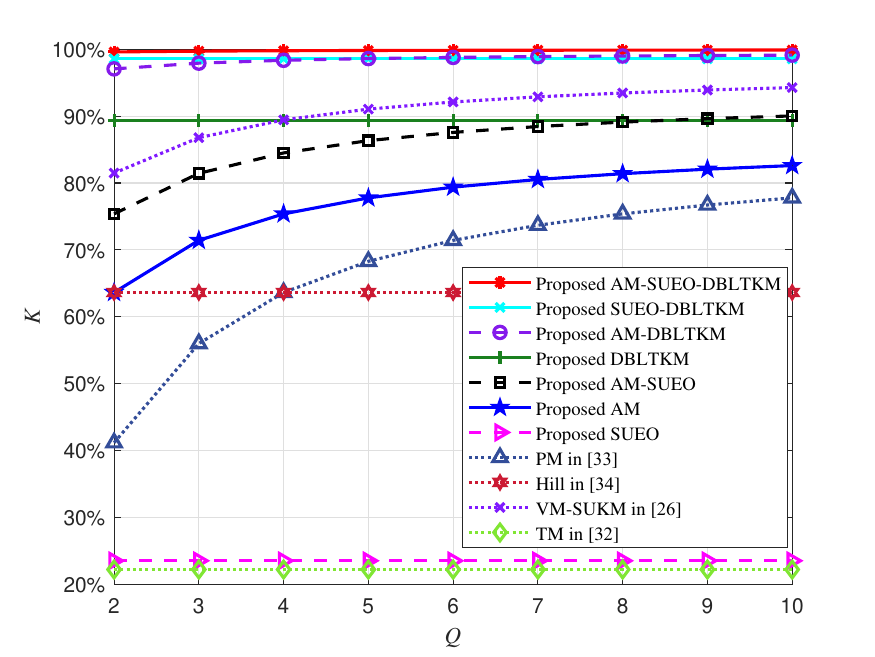}
	\caption{Tag storage saving ratio $K$ versus the total number of variable modulus $Q$ when the number of key matrices $N=2$ and the plaintext length $n=2$.}
    \label{Nc}
\end{figure}

Fig.~\ref{Nc} depicts $K$ versus $Q$ when $N=2$ and $n=2$. 
Among the seven proposed encryption algorithms, it can be seen that except for $K_{\text{SUEO}}$, $K_{\text{DBLTKM}}$ and $K_{\text{SUEO-DBLTKM}}$, which are independent of $Q$ and have always maintained a constant tag storage saving ratio $K$, the $K$ of the other four algorithms increases with the increase of $Q$.
When $Q=3$, the descending order of $K$ is: $K_{\text{AM-SUEO-DBLTKM}}>K_{\text{SUEO-DBLTKM}}>K_{\text{AM-DBLTKM}}
>K_{\text{ DBLTKM}}>K_{\text{AM-SUEO}}>K_{\text{AM}}>K_{\text{ SUEO}}$. In addition, the proposed seven algorithms, and the other four existing algorithms, i.e., the VM-SUKM algorithm in \cite{WANGEfficient}, the TM algorithm in \cite{ChenGeneration}, the PM algorithm in \cite{MansourALMS}, and the Hill algorithm in \cite{FengCryptanalysis}, are compared and analyzed. Similarly, the TM algorithm and the Hill algorithm do not consider updating the modulus, so their $K$ does not change with the change of $Q$. As shown in Fig.~\ref{Nc}, regardless of how $Q$ increases, the proposed AM-SUEO-DBLTKM algorithm, SUEO-DBLTKM algorithm, and AM-DBLTKM algorithm all have better tag storage saving ratio than existing algorithms.
\begin{figure}[h]
\centering
	\includegraphics [width=0.9\textwidth]{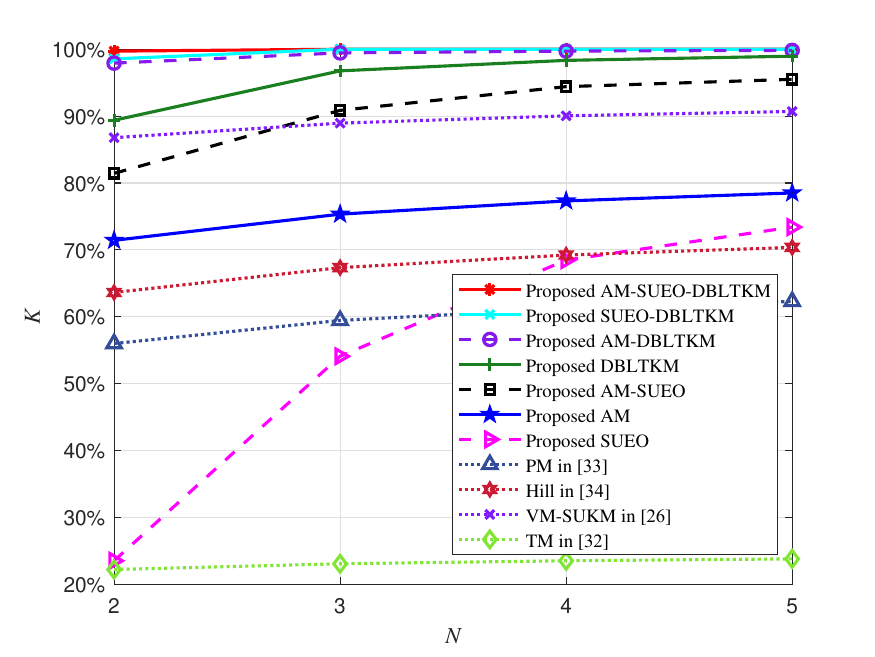}
	\caption{Tag storage saving ratio $K$ versus the number of key matrices $N$ when the plaintext length $n=2$ and the total number of variable modulus $Q=3$.}
    \label{KN}
\end{figure}

Fig.~\ref{KN} illustrates $K$ versus $N$ when $n=2$ and $Q=3$. 
From Fig.~\ref{KN}, it can be concluded that as $N$ increases, the $K$ of all other proposed seven algorithms increases with the increase of $N$.
Due to the fact that the SUEO algorithm mainly considers updating the multiplication order of the key matrix, which is highly correlated with the number of key matrices, its $K$ increases more obviously with the increase of $N$.
Moreover, the descending order of $K$ at $N=2$ in Fig.~\ref{KN} is consistent with the analysis of $Q=3$ in Fig.~\ref{Nc}. 
When $N \geq 3$, the tag storage savings of the proposed AM-SUEO-DBLTKM algorithm, SUEO-DBLTKM algorithm, AM-DBLTKM algorithm and DBLTKM algorithm are much higher than those of the other four existing algorithms \cite{WANGEfficient}, \cite{ChenGeneration}, \cite{MansourALMS}, \cite{FengCryptanalysis}.
In the application scenario where multiple key matrices need to be stored, almost 100\% of the tag storage space optimization can be achieved by using the proposed AM-SUEO-DBLTKM algorithm, SUEO-DBLTKM algorithm, and AM-DBLTKM algorithm. In other words, these three algorithms can significantly save the storage space of low-cost RFID tags.

\begin{figure}[h]
\centering
	\includegraphics [width=0.9\textwidth]{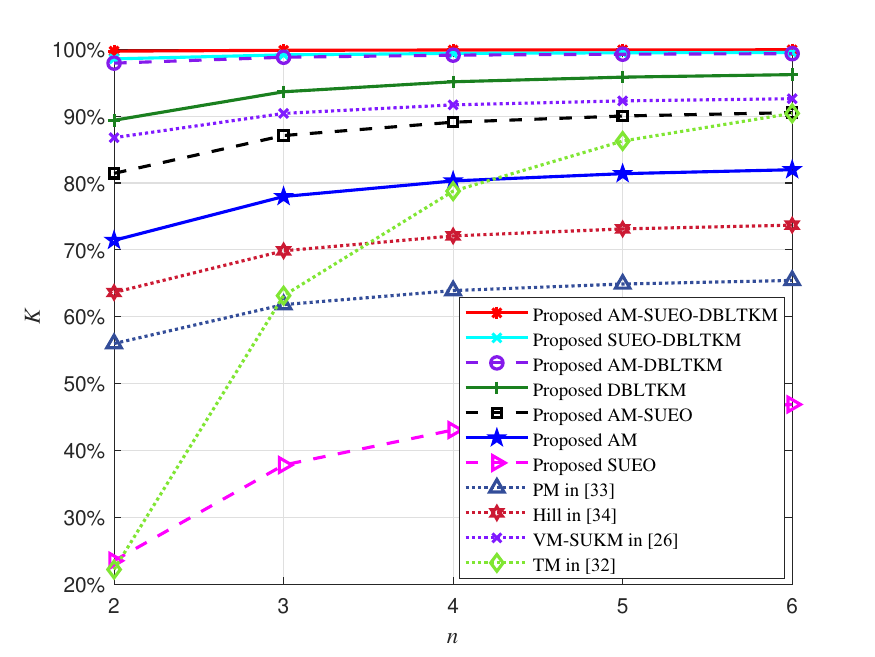}
	\caption{Tag storage saving ratio $K$ versus the plaintext length $n$ when the number of key matrices $N=2$ and the total number of variable modulus $Q=3$.}
    \label{Kn}
\end{figure}

Fig.~\ref{Kn} describes $K$ versus $n$ when $N=2$ and $Q=3$. 
The $K$ of the proposed algorithm increases with the increase of $n$, and finally the steady state is obtained.
Obviously, for plaintext of different lengths, the proposed encryption algorithm can significantly reduce the storage space required by the tag to varying degrees.
Furthermore, when $n\geq2$, the descending order of $K$ is consistent with the analysis results of $Q=3$ in Fig.~\ref{Nc} and $N=2$ in Fig.~\ref{KN}.
When $n\leq6$, regardless of how quickly the tag storage saving of the TM algorithm in \cite{ChenGeneration} increase with plaintext length, the proposed AM-SUEO-DBLTKM algorithm, SUEO-DBLTKM algorithm, AM-DBLTKM algorithm, and DBLTKM algorithm always outperform the other four existing algorithms.

\begin{table}[h]
    \centering
    \footnotesize
    \renewcommand{\arraystretch}{1.0}
    \setlength{\tabcolsep}{7.5pt}
    \caption{Tag storage saving ratio $K$ of the proposed seven encryption algorithm for different plaintext length $n$ and the number of key matrices $N$ when the total number of variable modulus $Q=2$.}
    \label{tab:DBKM-SUEO-SUM}
    \scalebox{1}{
    \begin{tabular}{cccc}
    \toprule
      ~  &\textbf{\makecell[c]{$n=2$\\ $N=2$}} &\textbf{\makecell[c]{$n=2$ \\ $N=3$}}&\textbf{\makecell[c]{$n=3$\\ $N=3$}} \\ \midrule    
      Proposed SUEO  & $23.53\%$ & $54.10\%$ & $68.38\%$  \\
      Proposed AM & $63.64\%$ & $67.35\%$ & $71.56\%$ \\
      Proposed AM-SUEO  & $75.38\%$ & $87.14\%$ & $91.50\%$ \\
      Proposed DBLTKM  & $89.38\%$ & $96.76\%$ & $98.48\%$ \\
      Proposed AM-DBLTKM  & $97.06\%$ & $99.18\%$ & $99.62\%$  \\
     Proposed SUEO-DBLTKM  & $98.56\%$ & $99.99\%$ & $100.00\%$  \\
      Proposed AM-SUEO-DBLTKM  & $99.59\%$ & $100.00\%$ & $100.00\%$  \\
    \bottomrule
    \end{tabular}
    }
\end{table}

Based on Fig.~\ref{Nc}, Fig.~\ref{KN}, and Fig.~\ref{Kn}, regardless of the values of $Q$, $N$, and $n$, the proposed AM-SUEO-DBLTKM algorithm, SUEO-DBLTKM algorithm and AM-DBLTKM algorithm almost achieve the optimal value of $K$. In order to confirm the advantages of these three algorithms more clearly and intuitively, the $K$ values of the proposed seven algorithms with different values of $n$ and $N$ when $Q=2$ are listed in Table \ref{tab:DBKM-SUEO-SUM}. It can be seen that the proposed joint AM-SUEO-DBLTKM encryption algorithm can save at least 99.59\% of the tag storage under the configuration of the plaintext length $n=2$, the number of key matrices $N=2$, and the total number $Q=2$ of integer divisor of modulus $p$. Based on Table \ref{tab:DBKM-SUEO-SUM}, for the diverse security requirements in different application scenarios, suitable encryption algorithms can be selected for deployment to ensure a perfect balance between the high security of the algorithms and the low cost of the tags.

\begin{figure}[h]
\centering
	\includegraphics [width=0.9\textwidth]{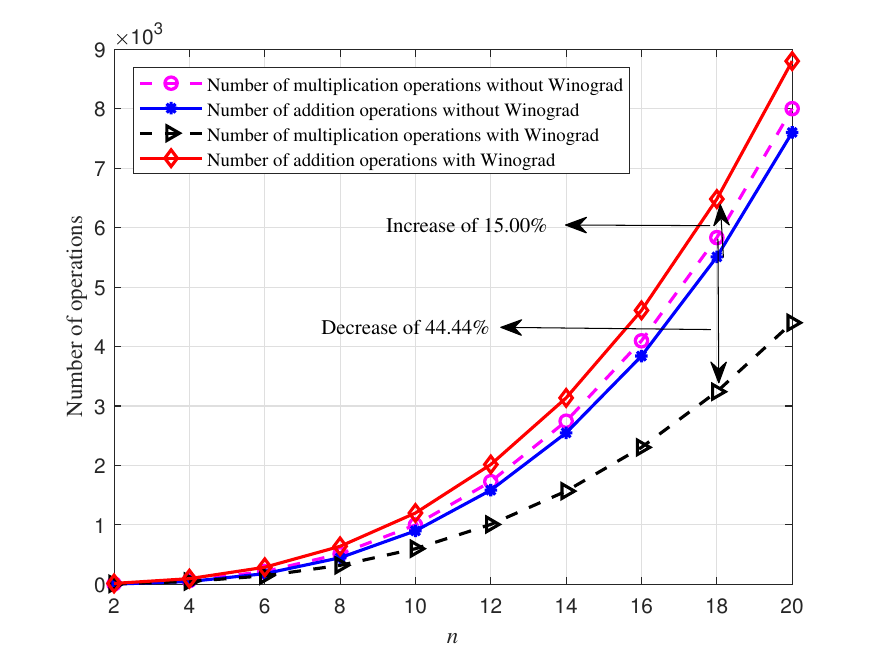}
	\caption{The number of multiplication or addition operations involved in encryption versus the plaintext length $n$.}
    \label{Winograd}
\end{figure}

In order to illustrate more clearly the acceleration design of the Winograd algorithm for key matrix encryption algorithms, the number of multiplication or addition operations involved in encryption versus the plaintext length $n$ is depicted in Fig.~\ref{Winograd}.
When $n=18$, utilizing the Winograd algorithm, addition operations only increased by 15.00\%, while complex multiplication operations decreased by 44.44\%. The significant reduction in complex multiplication operations is very beneficial for low-cost RFID tags. 
Therefore, introducing the fast convolutional Winograd algorithm is very beneficial for improving the real-time performance of protocol authentication process.

\section{Conclusion}
In this paper, in order to solve the problem that it is difficult to balance the security and tag storage costs in the traditional key matrix encryption algorithm, the AM encryption algorithm was proposed from the perspective of update modulus, the SUEO encryption algorithm was presented from the perspective of update encryption order, and the DBLTKM encryption algorithm was designed from the perspective of updating the key matrix. Subsequently, a two-way RFID authentication protocol was constructed based on the joint algorithm of the three algorithms, namely the AM-SUEO-DBLTKM algorithm. Through BAN logic and security analysis, it was demonstrated that the proposed AM-SUEO-DBLTKM-RFID protocol can effectively satisfy the various security requirements in mobile RFID systems.
With a configuration when 2 key matrices, the plaintext length is $2$ and the total number of integer divisor of modulus $p$ is $2$, the joint AM-SUEO-DBLTKM algorithm can save $99.59\%$ of tag storage and is highly suitable for systems with low-cost RFID tags.


\appendices
\section{The proof of Corollary 2}

Since $\mathbf{A}\times \mathbf{B}\equiv \mathbf{E} \bmod (p)$, then $\mathbf{A}\times \mathbf{B}\equiv p\mathbf{C}+\mathbf{E}$, where $\mathbf{C}$ is an arbitrary integer matrix. And because $q$ is the integer divisor of $p$, $p=aq$, where $a$ is a positive integer. Then $\mathbf{A}\times \mathbf{B}\equiv q\mathbf{D}+\mathbf{E}\equiv aq\mathbf{C}+\mathbf{E}$, where $\mathbf{D}$ is an arbitrary integer matrix. In summary, it is proven that $\mathbf{A}\times \mathbf{B}\equiv \mathbf{E} \bmod (q)$.

Hence, the proof is completed.

\section{The proof of Corollary 3}

According to (\ref{keymatrixE}), it can be inferred from $\mathbf{A}_1 \times \mathbf{t} \mod (p)=\mathbf{c}_1$ that $\det(\mathbf{A}_1)$ and modulus $p$ are mutually prime. Similarly, it can be inferred from $\mathbf{A}_2 \times \mathbf{c}_1 \mod (p)=\mathbf{c}_2$ that $\det(\mathbf{A}_2)$ and modulus $p$ are mutually prime. According to \textbf{\textit{Lemma 3}}, it can be seen that $\det(\mathbf{A}_2)\times \det(\mathbf{A}_1)$ and the modulus $p$ are mutually prime. According to \textbf{\textit{Lemma 4}}, it can be seen that $\det(\mathbf{A}_2\times \mathbf{A}_1)
=\det(\mathbf{A}_2)\times \det(\mathbf{A}_1)$. Therefore, there exists a matrix that is the modulus $p$-inverse of matrix $\mathbf{A}_2\times \mathbf{A}_1$. Similarly, there exists a matrix that is the modulus $p$-inverse of matrix $\mathbf{A}_1\times \mathbf{A}_2$. According to \textbf{\textit{Lemma 5}}, it can be seen that $\mathbf{A}_2\times \mathbf{A}_1\neq\mathbf{A}_1\times \mathbf{A}_2$, then $\mathbf{A}_2\times \mathbf{A}_1\times \mathbf{t} \mod (p)\neq\mathbf{A}_1\times \mathbf{A}_2\times \mathbf{t} \mod (p)$, namely $\mathbf{c}_2\neq\mathbf{c}_4$.

Hence, the proof is completed.

\section{The proof of Corollary 4}
According to (\ref{keymatrixE}), it can be inferred from $\mathbf{A} \times \mathbf{t} \mod (p)=\mathbf{c}_1$ that $\det(\mathbf{A})$ and modulus $p$ are mutually prime. According to \textbf{\textit{Lemma 7}}, it can be seen that $\det(\mathbf{A}^T)$ and the modulus $p$ are mutually prime. Therefore, there exists $\mathbf{A}^T \times \mathbf{t} \mod (p)=\mathbf{c}_2$. 

Hence, the proof is completed.

\section{The proof of Corollary 5}
According to (\ref{keymatrixE}), it can be inferred from $\mathbf{A}_1 \times \mathbf{t}_1 \mod (p)=\mathbf{c}_1$ that $\det(\mathbf{A}_1)$ and modulus $p$ are mutually prime. Similarly, it can be inferred from $\mathbf{A}_2 \times \mathbf{t}_2 \mod (p)=\mathbf{c}_2$ that $\det(\mathbf{A}_2)$ and modulus $p$ are mutually prime. According to \textbf{\textit{Lemma 3}}, it can be seen that $\det(\mathbf{A}_1)\times \det(\mathbf{A}_2)$ and the modulus $p$ are mutually prime. According to \textbf{\textit{Lemma 6}}, it can be seen that $\det\left(\begin{array}{cc}\mathbf{A}_1&\\&\mathbf{A}_2\end{array}\right)
=\det(\mathbf{A}_1)\times \det(\mathbf{A}_2)$. Therefore, there exists a matrix that is the modulus $p$-inverse of block matrix $\left(\begin{array}{cc}\mathbf{A}_1&\\&\mathbf{A}_2\end{array}\right)$. 

Hence, the proof is completed.

%


\begin{thebibliography}{1}

\bibitem{6G}
R. Liu, H. Lin, H. Lee, F. Chaves, H. Lim and J. Sköld, ``Beginning of the Journey Toward 6G: Vision and Framework,'' \emph{IEEE Commun. Mag.}, vol. 61, no. 10, pp. 8-9, Oct. 2023.

\bibitem{KIoT-QWatch}
K. Fizza, P. P. Jayaraman, A. Banerjee, N. Auluck and R. Ranjan, ``IoT-QWatch: A Novel Framework to Support the Development of Quality-Aware Autonomic IoT Applications,'' \emph{IEEE Internet Things J.}, vol. 10, no. 20, pp. 17666-17679, Oct. 2023.

\bibitem{HNew}
H. F. Hammad, ``New Technique for Segmenting RFID Bandwidth for IoT Applications,'' \emph{IEEE J. Radio Freq. Identif.}, vol. 5, no. 4, pp. 446-450, Dec. 2021.


\bibitem{FBroadband/Dual-Band}
F. Erman, S. Koziel and L. Leifsson, ``Broadband/Dual-Band Metal-Mountable UHF RFID Tag Antennas: A Systematic Review, Taxonomy Analysis, Standards of Seamless RFID System Operation, Supporting IoT Implementations, Recommendations, and Future Directions,'' \emph{IEEE Internet Things J.}, vol. 10, no. 16, pp. 14780-14797, Aug. 2023.


\bibitem{GPhysical-Layer}
G. Khadka, B. Ray, N. C. Karmakar and J. Choi, ``Physical-Layer Detection and Security of Printed Chipless RFID Tag for Internet of Things Applications,'' \emph{IEEE Internet Things J.}, vol. 9, no. 17, pp. 15714-15724, Sept. 2022.

\bibitem{FPUF}
F. L. Ţiplea and C. Hristea, ``PUF Protected Variables: A Solution to RFID Security and Privacy Under Corruption With Temporary State Disclosure,'' \emph{IEEE Trans. Inf. Forensics Secur.}, vol. 16, pp. 999-1013, 2021.

\bibitem{SundaresanA}
S. Sundaresan, R. Doss, S. Piramuthu and W. Zhou, ``A Robust Grouping Proof Protocol for RFID EPC C1G2 Tags,''\emph{IEEE Trans. Inf. Forensics Secur.}, vol. 9, no. 6, pp. 961-975, Jun. 2014.


\bibitem{GopeLightweight}
P. Gope, J. Lee and T. Q. S. Quek, ``Lightweight and Practical Anonymous Authentication Protocol for RFID Systems Using Physically Unclonable Functions,'' \emph{IEEE Trans. Inf. Forensics Secur.}, vol. 13, no. 11, pp. 2831-2843, Nov. 2018.


\bibitem{ZhangDevice-Side}
Q. Zhang, X. Zhou, H. Zhong, J. Cui, J. Li and D. He, ``Device-Side Lightweight Mutual Authentication and Key Agreement Scheme Based on Chameleon Hashing for Industrial Internet of Things,'' \emph{IEEE Trans. Inf. Forensics Secur.}, vol. 19, pp. 7895-7907, 2024.


\bibitem{QiSAE}
M. Qi, W. Hu and Y. Tai, ``SAE+: One-Round Provably Secure Asymmetric SAE Protocol for Client-Server Model,'' \emph{IEEE Trans. Inf. Forensics Secur.}, vol. 19, pp. 3906-3913, 2024.



\bibitem{DAn}
D. He and S. Zeadally, ``An Analysis of RFID Authentication Schemes for Internet of Things in Healthcare Environment Using Elliptic Curve Cryptography,'' \emph{IEEE Internet Things J.}, vol. 2, no. 1, pp. 72-83, Feb. 2015.

\bibitem{KLightweight}
K. Fan, W. Jiang, H. Li and Y. Yang, ``Lightweight RFID Protocol for Medical Privacy Protection in IoT,'' \emph{IEEE Trans. Industr. Inform.}, vol. 14, no. 4, pp. 1656-1665, Apr. 2018.

\bibitem{CPrivacy}
C. Hristea and F. L. Ţiplea, ``Privacy of Stateful RFID Systems With Constant Tag Identifiers,'' \emph{IEEE Trans. Inf. Forensics Secur.}, vol. 15, pp. 1920-1934, 2020. 

\bibitem{HAnalysis}
H. A. Abdulghani, N. A. Nijdam and D. Konstantas, ``Analysis on Security and Privacy Guidelines: RFID-Based IoT Applications,'' \emph{IEEE Access}, vol. 10, pp. 131528-131554, 2022.

\bibitem{SA}
S. Qiu, G. Xu, H. Ahmad and L. Wang, ``A Robust Mutual Authentication Scheme Based on Elliptic Curve Cryptography for Telecare Medical Information Systems,'' \emph{IEEE Access}, vol. 6, pp. 7452-7463, 2018.

\bibitem{KCloud-based}
K. Fan, Q. Luo, K. Zhang, and Y. Yang, ``Cloud-based lightweight secure RFID mutual authentication protocol in IoT,'' \emph{Inf. Sci.}, vol. 527, pp. 329–340, Jul. 2020.

\bibitem{VRseap}
V. Kumar, M. Ahmad, et al., ``Rseap: Rfid based secure
and efficient authentication protocol for vehicular cloud
computing,'' \emph{Veh. Commun.}, vol. 22, 2020, p.
100213.

\bibitem{MRseap2}
M. Safkhani, C. Camara, et al., ``Rseap2: An enhanced
version of rseap, an rfid based authentication protocol for
vehicular cloud computing,'' \emph{Veh. Commun.},
vol. 28, 2021, p. 100311.

\bibitem{Yang2022Delegating}
A. Yang, J. Weng, K. Yang, C. Huang and X. Shen, ``Delegating Authentication to Edge: A Decentralized Authentication Architecture for Vehicular Networks,'' \emph{IEEE Trans. Intell. Transp. Syst.}, vol. 23, no. 2, pp. 1284-1298, Feb. 2022.

\bibitem{KFanLightweight}
K. Fan, Y. Gong, C. Liang, H. Li, and Y. Yang. ``Lightweight and ultralightweight RFID mutual authentication protocol with cache in the reader for IoT in 5G.'' \emph{Secur. Commun. Netw.}, vol. 9, no. 16, pp. 3095-3104, 2016.

\bibitem{CTTowards}
C. T. Li, C. C. Lee, C. Y. Weng, and C. M. Chen, ``Towards secure authenticating of cache in the reader for RFID-based IoT systems,'' \emph{Peer-to-Peer Netw. Appl.}, vol. 11, no. 1, pp. 198–208, 2018.

\bibitem{SDesigning}
S. Jangirala, A. K. Das and A. V. Vasilakos, ``Designing Secure Lightweight Blockchain-Enabled RFID-Based Authentication Protocol for Supply Chains in 5G Mobile Edge Computing Environment,'' \emph{IEEE Trans. Industr. Inform.}, vol. 16, no. 11, pp. 7081-7093, Nov. 2020.

\bibitem{SDrone}
S. O. Ajakwe, D. -S. Kim and J. -M. Lee, ``Drone Transportation System: Systematic Review of Security Dynamics for Smart Mobility,'' \emph{IEEE Internet Things J.}, vol. 10, no. 16, pp. 14462-14482, Aug. 2023.


%
%
%
%
%
%
%
%





\bibitem{FanPermutation}
Fan K, Kang J, Zhu S, Li H, Yang Y. ``Permutation Matrix Encryption Based Ultralightweight Secure RFID Scheme in Internet of Vehicles.'' \emph{Sensors.} 2019; 19(1):152.

\bibitem{YLuo}
Y. Luo, K. Fan, X. Wang, H. Li and Y. Yang, ``RUAP: Random rearrangement block matrix-based ultra-lightweight RFID authentication protocol for end-edge-cloud collaborative environment,'' \emph{China Commun.}, vol. 19, no. 7, pp. 197--213, Jul. 2022.

\bibitem{WANGEfficient}
Y Wang, X Lei, T Gao. ``Efficient RFID security authentication protocol based on variable modulus and self-updating key matrix''. \emph{Journal of Cryptologic Research}, 2022, 9(2): 210--222.

\bibitem{DSWMWinograd}
D. Wu, X. Fan, W. Cao and L. Wang, ``SWM: A High-Performance Sparse-Winograd Matrix Multiplication CNN Accelerator,'' \emph{ IEEE Trans. Very Large Scale Integr. VLSI Syst.}, vol. 29, no. 5, pp. 936-949, May 2021.



\bibitem{ShihabLightweight}
S. Shihab and R. AlTawy, ``Lightweight Authentication Scheme for Healthcare With Robustness to Desynchronization Attacks,'' \emph{IEEE Internet Things J.}, vol. 10, no. 20, pp. 18140-18153, Oct. 2023.

\bibitem{SABAN}
S. Dhaka, Y. -J. Chen, S. De and L. -C. Wang, ``A Lightweight Stochastic Blockchain for IoT Data Integrity in Wireless Channels,'' \emph{IEEE OJVT}, vol. 4, pp. 765-781, 2023.

\bibitem{ZhangIoT2023a}
Q. Zhang, Z. Wang, B. Wu, and G. Gui, ``A robust and practical solution to ADS-B security against denial-of-service attacks,'' \emph{IEEE Internet Things J.}, vol. 11, no. 8, pp. 13647--13659, Apr. 2024.


\bibitem{LA}
L. Chen, Y. Wang, T. Tu and M. Zhao, ``A PUF-based Security Authentication Protocol against Desynchronization Attacks,'' in \emph{2021 IEEE 6th International Conference on Signal and Image Processing (ICSIP)}, Nanjing, China, 2021, pp. 999-1003.

\bibitem{ChenGeneration}
Y, Chen, R, Xie, H, Zhang, et al. ``Generation of high-order random key matrix for Hill Cipher encryption using the modular multiplicative inverse of triangular matrices.'' \emph{Wireless Netw.}, (2023).

\bibitem{MansourALMS}
A. Mansour, K. M. Malik, A. Alkaff and H. Kanaan, ``ALMS: Asymmetric Lightweight Centralized Group Key Management Protocol for VANETs,'' \emph{IEEE Trans. Intell. Transp. Syst.}, vol. 22, no. 3, pp. 1663--1678, Mar. 2021.

\bibitem{FengCryptanalysis}
W. Feng, Z. Qin, J. Zhang and M. Ahmad, ``Cryptanalysis and Improvement of the Image Encryption Scheme Based on Feistel Network and Dynamic DNA Encoding,'' \emph{IEEE Access}, vol. 9, pp. 145459--145470, 2021.

















\end{thebibliography}
\end{document}